\documentclass{article}

\usepackage{arxiv}

\usepackage[utf8]{inputenc} % allow utf-8 input
\usepackage[T1]{fontenc}    % use 8-bit T1 fonts
\usepackage{hyperref}       % hyperlinks
\usepackage{url}            % simple URL typesetting
\usepackage{booktabs}       % professional-quality tables
\usepackage{amsfonts}       % blackboard math symbols
\usepackage{nicefrac}       % compact symbols for 1/2, etc.
\usepackage{microtype}      % microtypography
\usepackage{lipsum}		% Can be removed after putting your text content\usepackage{osameet3}
\usepackage{graphicx}
\usepackage{natbib}
\usepackage{doi}

\usepackage{indentfirst}
\usepackage{amsmath}
\usepackage{float}
\usepackage{setspace}

\title{Physics-informed Neural Network for Nonlinear Dynamics in Fiber Optics}

\date{} 					% Or removing it

\author{Xiaotian Jiang$^{1}$, Danshi Wang$^{1,*}$, Qirui Fan$^{2}$, Min Zhang$^{1}$, Chao Lu$^{3}$, and Alan Pak Tao Lau$^{2}$\\
\\
\emph{$^{1}$State Key Laboratory of Information Photonics and Optical Communications, Beijing University of Posts and}\\ \emph{Telecommunications, Beijing, 100876, China}\\
\emph{$^{2}$Photonics Research Centre, Department of Electrical Engineering, The Hong Kong Polytechnic University,} \\
\emph{Hung Hom, Kowloon, Hong Kong} \\
\emph{$^{3}$Photonics Research Centre, Department of Electronic and Information Engineering, The Hong Kong Polytechnic} \\
\emph{University, Hung Hom, Kowloon, Hong Kong} \\ 
\\
\emph{Email:$^{*}$danshi\_wang@bupt.edu.cn}
}
\renewcommand{\headeright}{}
\renewcommand{\undertitle}{}
\renewcommand{\shorttitle}{Physics-informed Neural Network for Nonlinear Dynamics in Fiber Optics}

\hypersetup{
pdftitle={Physics-informed Neural Network for Nonlinear Dynamics in Fiber Optics},
pdfauthor={Xiaotian Jiang, Danshi Wang, Qirui Fan, Min Zhang, Chao Lu, and Alan Pak Tao Lau},
pdfkeywords={Physics-informed neural network (PINN), Automatic differentiation, Fiber optics, Nonlinear dynamics, Nonlinear Schrödinger equation (NLSE)},
}

\begin{document}

\maketitle
\begin{spacing}{2.0}
\setlength{\parskip}{0em}
\begin{abstract}
	A physics-informed neural network (PINN) that combines deep learning with physics is studied to solve the nonlinear Schrödinger equation for learning nonlinear dynamics in fiber optics. We carry out a systematic investigation and comprehensive verification on PINN for multiple physical effects in optical fibers, including dispersion, self-phase modulation, and higher-order nonlinear effects. Moreover, both special case (soliton propagation) and general case (multi-pulse propagation) are investigated and realized with PINN. In the previous studies, the PINN was mainly effective for single scenario. To overcome this problem, the physical parameters (pulse peak power and amplitudes of sub-pulses) are hereby embedded as additional input parameter controllers, which allow PINN to learn the physical constraints of different scenarios and perform good generalizability. Furthermore, PINN exhibits better performance than the data-driven neural network using much less data, and its computational complexity (in terms of number of multiplications) is much lower than that of the split-step Fourier method. The results report here show that the PINN is not only an effective partial differential equation solver, but also a prospective technique to advance the scientific computing and automatic modeling in fiber optics.
\end{abstract}

% keywords can be removed

\section{Introduction}

The dynamics of nonlinear systems generally must be described by nonlinear models, whose theory is technically difficult and the detailed analytical solutions are always not obtainable.[1] To understand some aspects of such nonlinear dynamics, various complementary ideas and methods from many different fields of mathematics are of crucial importance.[2] Moreover, as is often the case for a relatively fast growing area of research, modeling and forecasting the dynamics of multiphysics and multiscale nonlinear systems remains an open scientific problem, including but not limited to physical, chemical, or biological systems. Therefore, the more extensive study, deeper analysis, and more efficient modeling on the complex nonlinear dynamical systems are particularly important. To well explain and characterize the nonlinear dynamics, decades of extensive explorations have made tremendous progress in diverse fields, such as the nonlinear vibration problem in nonlinear mechanics,[3] the three-body problem in celestial mechanics,[4] and the atmospheric turbulence problem in meteorology[5] by solving complex partial differential equations (PDEs) in these problems. Generally, these PDEs have no analytical solutions, and the great progress of numerical methods such as finite-difference method (FDM), finite elements method, spectral and even meshless methods has made it possible to obtain the approximate solutions. Despite great progress, these conventional numerical methods still have difficulties in solving physical problems with missing initial and boundary conditions (e.g., inferring the initial pulse from the received pulse). For solving inverse problems (e.g., inferring the physical parameters of PDE from limited data), numerical methods are often prohibitively expensive or even currently impossible. In addition, strict mathematical theories and expert knowledge are required for application of these techniques. 

In fiber optics, the nonlinear dynamics is the physical basis of fiber-based optical devices, optical information processing, photonic material design and optical signal transmission, which has greatly promoted the development of fiber lasers, fiber amplifiers, fiber waveguides, and fiber-optic communications.[6-10] For a fully understanding of the nonlinear dynamics in optical fibers, the propagation dynamics of optical pulses through a fiber can be governed by a fundamental nonlinear PDE, i.e., nonlinear Schrödinger equation (NLSE),[11] which cannot be solved analytically when dispersion and nonlinearity coexist. To solve NLSE, the split-step Fourier method (SSFM) and its modifications have been applied extensively for studying various nonlinear effects in optical fibers, mainly due to its straightforward implementation and relative accuracy compared with other numerical methods.[12,13] However, for long-distance scenarios with high nonlinearity, the step lengths of SSFM must be reduced considerably to meet the accuracy requirements which increases the computational complexity. Moreover, when NLSE is solved for solving high-dimensional problems (e.g., simulating wavelength-division-multiplexed (WDM) systems), the temporal resolution should be a small fraction of the entire bandwidth, resulting in the extremely massive mesh points in time domain and thus becoming a quite time-consuming process. Hence, it is desirable to develop more powerful and advanced techniques to directly solve the NLSE without complex numerical calculating.

In the past decade, deep learning (DL) is evolving rapidly and studied extensively to solve all sorts of issues, including various nonlinear dynamic systems.[14,15] At the beginning, DL was mainly applied as a black-box tool based on the data-driven idea without regard to any prior knowledge. Provided the input-output data pairs are available for training, this black-box method for data fitting learns the input-output relationship of an analytically intractable system in a data-driven manner. To date, various nonlinear problems in optical fibers have been solved by DLs with different neural network structures based on the massive data collected from either experiment or simulation, bridging the nonlinear mapping between the input parameters and output targets.[16,17,18,19] However, although this black-box approach can reduce the computational complexity to a certain extent and expand the types of problems that can be solved, its performance is heavily reliant on the quantity and the quality of data. Moreover, data-driven networks are usually trained without considering the underlying mathematics and physics, while such prior knowledge is essential for fully understanding the philosophy behind the nonlinear dynamics.

Recently, a physics-informed neural network (PINN) is proposed, [20,21] which fully considers the prior knowledge, including the essential mathematical equations, physical theories, and the corresponding constraint conditions of the target problems. Despite newly proposed, PINN has attracted widespread attention and been verified in multiple nonlinear dynamic systems governed by different PDEs, such as the Navier-Stokes equation of fluid dynamics,[20,22] the Burgers equation of aerodynamics,[20,23] and the heat transfer equation of thermodynamics.[24,25] Unlike data-driven neural networks that simply fit their outputs to the given inputs, PINN manifests itself as an automatic PDE solver that maps the independent variables of the PDE to its solutions. More specifically, the governing equations as well as physical constraints including the initial and boundary conditions constitute the network loss function and serve as the regularization mechanisms. Hence, PINN transforms the PDE solving problem into a loss function optimization problem. In this way, the network integrates the prior knowledge of mathematics and physics throughout the entire training process. To calculate the governing equations, automatic differentiation technique is employed,[26] and the differential terms of PDE are calculated automatically according to the auxiliary coordinates, without the exact data at these coordinates; hence, the data dependence of the model is greatly reduced. Since the complexity is mainly determined by the network structure, the computational complexity of PINN with stable structure is almost unchanged for different scenarios, which is quite different from numerical methods. Moreover, PINN has also achieved remarkable results in solving problems with missing physical constraints and inverse problems.[27,28] Therefore, PINN combines the advantages of both numerical and data-driven methods while overcoming the limitations of them. Currently, studies on nonlinear dynamics in fiber optics were mainly based on the numerical and data-driven methods, and only few PINN-based work was preliminarily in solving NLSE.[29,30] A systematic investigation and comprehensive verification on this subject is yet to be undertaken. Moreover, the current PINN models perform poor generalizability and cannot be directly applied when the physical constraints change without remodeling. Therefore, it is still necessary to separately establish models for different physical constraints in the study of nonlinear dynamics involving multiple scenarios in fiber optics, which greatly limits the application value of PINN.

In this study, we proposed a PINN-based solution to characterize the complex nonlinear dynamics and model multiple physical effects in fiber optics. The general context of our work is the recent development of the combination of DL and physics, which solves the complex PDEs with prior knowledge involving the governing equations and physical constraints of nonlinear dynamic systems. To improve the generalizability of PINN, the physical parameters (pulse peak power and amplitudes of sub-pulses) are embedded to the PINN input layer as physical parameter controllers, thereby allowing the PINN model to learn the different physical constraints and simulate the corresponding physical scenarios. We carried out the systematic investigation and comprehensive verification on multiple physical effects in optical fibers, including chromatic dispersion, self-phase modulation (SPM), and higher-order nonlinear effects. Moreover, both sepcial case (soliton propagation) and general case (multi-pulse propagation) were investigated and realized with PINN. the accuracy and the complexity of the PINN model are also studied in this work and results show that PINN exhibits better performance than data-driven methods with much less data, and the computational complexity in terms of multiplications is usually two orders of magnitude fewer than that of SSFM. Therefore, PINN is not only equivalent to a complex operator for solution of PDEs with various physical constraints, but can also act as a supplemental tool to advance the theoretical understanding of nonlinear dynamics in fiber optics.

\section{Concept and Principles}
\label{sec:headings}

Multiple linear and nonlinear effects occur in optical fibers, causing various distortions during pulse propagation, which can be well characterized by NLSE. The NLSE is a complex PDE containing multiple higher order differential terms. To calculate these differential terms for a given accuracy, more iterations and smaller step size are usually required using numerical methods, which increases the complexity tremendously. Instead of directly calculating the differential terms, the conventional DL methods usually rely on large amounts of simulation or experimental data for modeling. Different from them, PINN not only takes both accuracy and complexity into account, but also makes full use of the prior knowledge of underlying mathematics and physics to provide a reliable and efficient solution for the complex nonlinear dynamic systems. Next, we will illustrate the theory of PINN in detail and explain its modeling principles of nonlinear dynamics in fiber optics.

\subsection{Overview of Physics-Informed Neural Network}
Most PDEs can be expressed in the following general form including governing equation, initial and boundary conditions:
\begin{equation}
\begin{split}
    &h_{t}+\mathcal{N}_{x}(h)=0,x\in{\Omega},t\in{[0,T]}\\
    &h(x,0)=h_0(x,0),x\in{\Omega},t=0\\
    &h(x,t)=h_b(x,t),x\in{\partial{\Omega}},t\in{[0,T]}
\end{split}
\end{equation}
where $h(x,t)$ denotes the solutions of the PDE, $h_t$ is the temporal derivative. $\mathcal{N}_{x}$ is a generalized nonlinear differential operator, including spatial derivatives and other nonlinear terms such as $|h|^2$. $\Omega$ and $\partial{\Omega}$ denote the computational domain and boundary. Further, $h_0 (x,t)$ and $h_b(x,t)$ denote the solutions at $t = 0$ and the boundary, respectively.

\begin{figure}
	\centering
	\includegraphics[width = .8\textwidth]{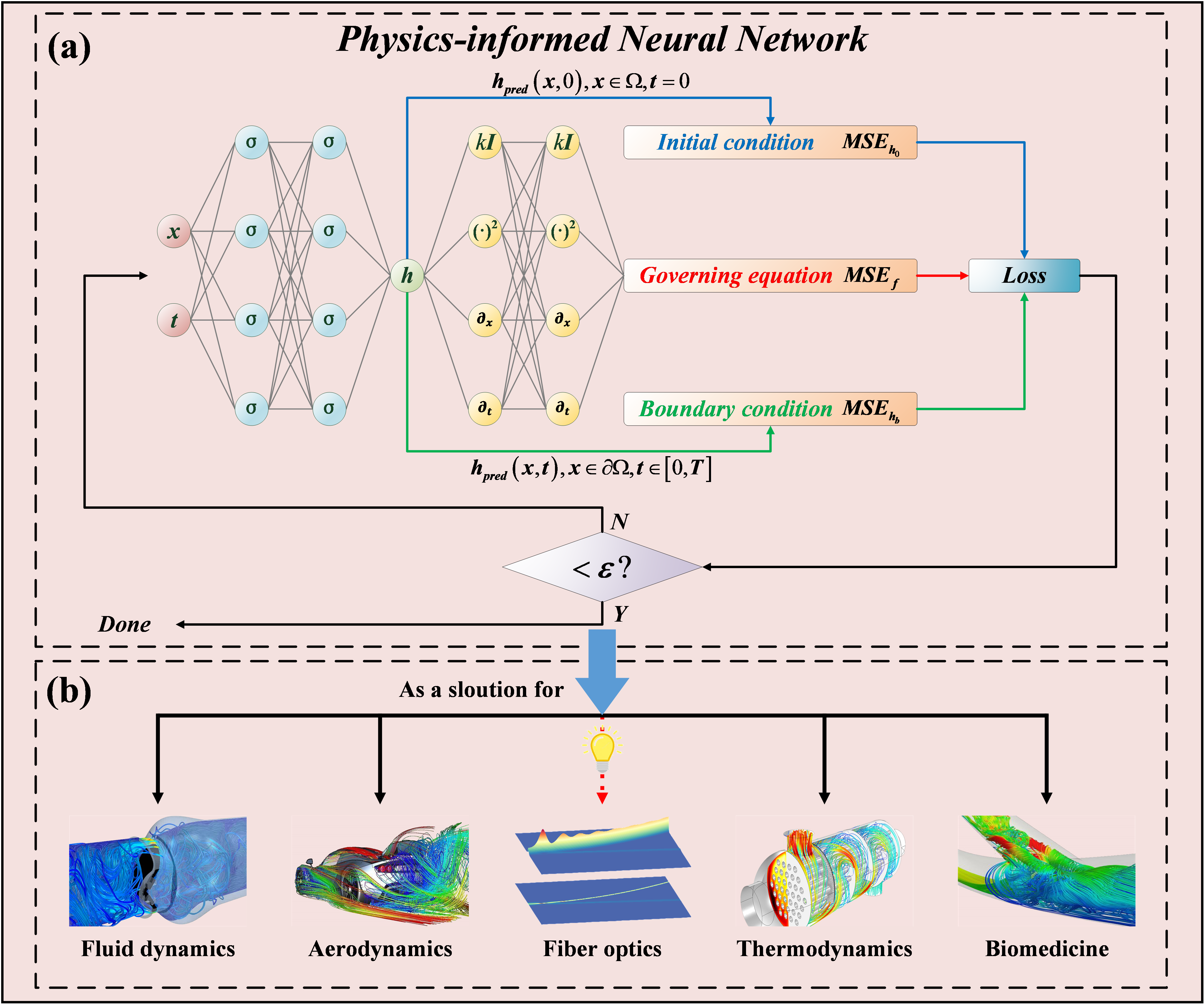}
	\setlength{\abovecaptionskip}{0pt}
    \setlength{\belowcaptionskip}{0pt}
	\caption{a) Principle of physics-informed neural network and b) applications of physics-informed neural network: fluid dynamics, aerodynamics, thermodynamics, biomedicine, or fiber optics.}
	\label{fig:fig1}
\end{figure}
\raggedbottom

To solve the PDE with initial and boundary conditions given in Equation (1), a PINN network composed of a multi-layer deep neural network is established to approximate $h$, as shown in Figure 1a. The PINN network takes the temporal-spatial auxiliary coordinate $(x,t)$ as input and exports the predicted solution $h(x,t)$ as output. To meet the constraints of the governing equation, it is necessary to accurately calculate the temporal derivatives, spatial derivatives, and the nonlinear terms, so as to estimate the value of the governing equation. With the help of automatic differentiation, the differential and nonlinear terms of $h$ can be calculated through equation parameters and input auxiliary coordinates. Moreover, as a mesh-free method, automatic differentiation does not suffer from errors such as truncation and round-off errors, which are quite common in conventional numerical methods. Furthermore, the solutions at $t = 0$ and the boundary will be predicted and compared with the exact data at these places to satisfy the initial and boundary conditions. Therefore, the network parameters can be learned by minimizing the mean square error (MSE) loss of the governing equation, initial and boundary conditions, which is expressed as
\begin{equation}
\begin{split}
    Loss&=MSE_f+MSE_{h_{0}}+MSE_{h_{b}}\\
    &=\frac{1}{N_f}\sum_{i=1}^{N_f}|h_{t}+\mathcal{N}_{x}(h)|^2+\frac{1}{N_0}\sum_{i=1}^{N_0}|h_{pred}(x^i,0)-h_0(x^i,0)|^2+\frac{1}{N_b}\sum_{i=1}^{N_b}|h_{pred}(x^i,t^i)-h_b(x^i,t^i)|^2
\end{split}
\end{equation}
where $MSE_f$, $MSE_{h_0}$, and $MSE_{h_b}$ penalize the residuals of the governing equation, initial and boundary conditions, respectively. $N_f$, $N_0$, and $N_b$ are the numbers of auxiliary coordinates required to calculate each MSE term. $h_{pred}(x^i,t^i)$ denotes the solution at $(x^i,t^i)$ predicted by PINN. Note that the boundary condition given in Equation (2) is the Dirichlet boundary condition and the boundary derivative should also be included when the PDE has a mixed boundary condition.

When the loss is minimized to a small value $\epsilon$ after multiple iterations, the PINN network can be equivalent to a complex operator that solves this PDE. As shown in Figure 1b, this advanced technology has achieved great success in fluid dynamics, aerodynamics, thermodynamics, and biomedicine, which greatly motivates us to explore if PINN is possible to fundamentally reassess the nonlinear dynamics in fiber optics from a fresh perspective.

\subsection{PINN-based Modeling for Nonlinear Dynamics in Fiber Optics}
For nonlinear dynamics in fiber optics, the NLSE can characterize various physical effects in optical fibers, such as group velocity dispersion (GVD), third-order dispersion (TOD), and self-phase modulation. By introducing dimensionless parameters, the NLSE can be simplified to the following normalized form:[11]
\begin{equation}
h_z+\frac{\alpha}{2}h+\frac{i}{2}\beta_2h_{tt}-\frac{1}{6}\beta_3h_{ttt}-iN^2[|h|^2h+is(|h|^2h)_t-\tau_R(|h|^2)_th]=0
\end{equation}
where the dimensionless parameters are defined as
\begin{equation}
    h=\frac{H}{\sqrt{P_0}},z=\frac{Z}{L_D},t=\frac{T}{T_0},N^2=\frac{L_D}{L_{NL}}=\frac{\gamma P_0T^2_0}{|\beta_2|^2},L_D=\frac{T^2_0}{|\beta_2|^2},L_{NL}=\frac{1}{\gamma P_0},s=\frac{1}{\omega_0T_0},\tau_R=\frac{T_R}{T_0}
\end{equation}
Here, $H(Z,T)$ is the complex envelope of a slowly varying optical field. $Z$ and $T$ are the propagation distance and time scale in a frame of reference moving with the pulse at the group velocity. The dimensionless parameters $h$, $z$, and $t$ are the normalized values of $H$, $Z$, and $T$ under peak power $P_0$, dispersion length $L_D$, and pulse width $T_0$. $N^2$ is the ratio of $L_D$ to the nonlinear length $L_{NL}$, which governs the relative impact of GVD and SPM effects on the pulse evolution along the fiber. The propagation parameters $\alpha$, $\beta_2$, $\beta_3$, and $\gamma$ reflect the power attenuation, GVD, TOD, and the strength of the nonlinear effects during the pulse evolution. For ultrashort optical pulses with $T_0 < 1$ ps, higher-order nonlinear effects such as self-steepening (SS)[31] and intrapulse Raman scattering (IRS),[32] as expressed by the last two terms of Equation (3), should be considered, where $s$ and $\tau_R$ are the normalized parameters of SS and IRS, $\omega_0$ and $T_R$ in Equation (4) are the central angular frequency of the pulse and the first moment of the Raman response function, respectively.

For optical fibers, the physical constraint is mainly the initial pulse, which directly influences the evolution characteristics. Since the NLSE is a complex-valued equation, it is advisable to divide this governing equation into real- and imaginary-part, which allows the PINN applicable to optical fibers. Thus, we write $h(z,t)$ in rectangular form as $h(z,t)=u(z,t)+iv(z,t)$, where $u(z,t)$ and $v(z,t)$ are the real and imaginary parts of $h(z,t)$. Accordingly, the NLSE in Equation (3) can be rewritten in separated real- and imaginary-part form:
\begin{equation}
    f(u,v)+ig(u,v)=0
\end{equation}
where $f$ and $g$ are the real- and imaginary-part functions of the NLSE, and are expressed as:
\begin{equation}
\begin{split}
    &f:u_z+\frac{\alpha}{2}u-\frac{1}{2}\beta_2v_{tt}-\frac{1}{6}\beta_3u_{ttt}+N^2\{(u^2+v^2)v+s[(u^2+v^2)u]_t-\tau_R(|u^2+v^2|)_tv\}\\
    &g:v_z+\frac{\alpha}{2}v-\frac{1}{2}\beta_2u_{tt}-\frac{1}{6}\beta_3v_{ttt}+N^2\{(u^2+v^2)u+s[(u^2+v^2)v]_t-\tau_R(|u^2+v^2|)_tu\}
\end{split}
\end{equation}

Considering both the governing equations and initial condition, the loss function of PINN for the nonlinear dynamics in fiber optics can be expressed as
\begin{equation}
\begin{split}
    Loss&=MSE_{u_{0}}+MSE_{v_{0}}+MSE_f+MSE_g\\
    &=\frac{1}{N_0}\sum_{i=1}^{N_0}|u_{pred}(0,t^i)-u_0(0,t^i)|^2+\frac{1}{N_0}\sum_{i=1}^{N_0}|v_{pred}(0,t^i)-v_b(0,t^i)|^2\\
    &+\frac{1}{N_f}\sum_{i=1}^{N_f}|f(z^i,t^i)|^2+\frac{1}{N_f}\sum_{i=1}^{N_f}|g(z^i,t^i)|^2
\end{split}
\end{equation}
where $MSE_{u_0}$ and $MSE_{v_0}$ correspond to the real- and imaginary-part constraints of the initial pulse, while $MSE_f$ and $MSE_g$ penalize the training results that do not satisfy the constraints of governing equations. Further, $u_0 (0,t^i)$ and $v_0 (0,t^i)$ denote the real and imaginary parts of the initial pulse, respectively; $u_{pred}(0,t^i)$ and $v_{pred}(0,t^i)$ are the predicted outputs of real and imaginary parts with input $(0,t^i)$; $f(z^i,t^i)$ and $g(z^i,t^i)$ are the results of Equation (6) calculated by PINN with input $(z^i,t^i)$. $N_0$ and $N_f$ are the numbers of auxiliary coordinates required to calculate the MSE terms of the initial conditions, and governing equations.
\begin{figure}[H]
	\centering
	\includegraphics[width = 1.0\textwidth]{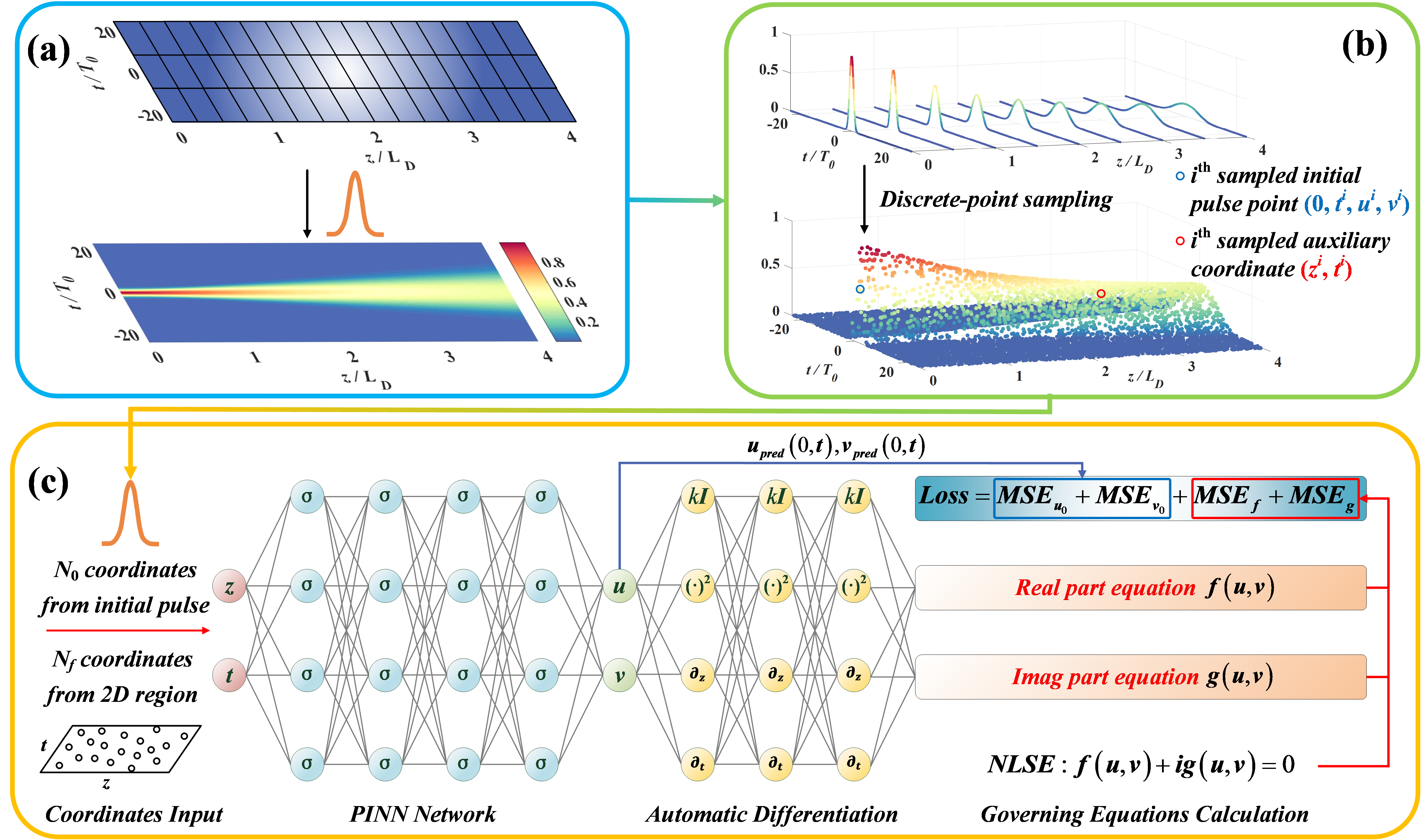}
	\setlength{\abovecaptionskip}{0pt}
    \setlength{\belowcaptionskip}{0pt}
	\caption{Process of modeling the nonlinear dynamics in fiber optics using PINN: a) temporal-spatial 2D region and the exact evolution data generated by SSFM as a reference, b) discrete-points sampling for PINN, c) the PINN structure constructed for solving the NLSE in fiber.}

\end{figure}
\raggedbottom

Figure 2 shows the detailed process of modeling the nonlinear dynamics in fiber optics using PINN. As shown in Figure 2a, a temporal-spatial 2D region is constructed to study the evolution dynamics of pulses in fiber, and the exact evolution data of various pulses can be generated by SSFM as a reference. To meet the constraints of initial condition and governing equations, $N_0$ data points of the initial pulse and $N_f$ auxiliary coordinates in the 2D region are sampled, as shown in Figure 2b. Note that the exact value at these auxiliary coordinates sampled is unknown. Following the PINN network and the automatic differentiation shown in Figure 2c, the temporal-spatial coordinates $(z,t)$ are used to calculate the four MSE terms in Equation (7) and the nonlinear dynamics in fiber optics can be well characterized after multiple iterations.

\section{Demonstrations and Results}
\label{sec:others}
As the modeling process mentioned above, a 2D region of $[-24T_0, 24T_0] \times [0, 4L_D]$ was constructed to reveal the pulse evolution. For typical standard single-mode fibers, $\beta_2$ is approximately -20 ps/km, $T_0$ is approximately 10-100 ps, and $L_D$ is approximately 5-500 km, according to Equation (4). $N_f=50,000$ auxiliary coordinates were sampled following the space-filling Latin hypercube sampling strategy[33] for calculating the $MSE_f$ and $MSE_g$ terms. The other two MSE terms which govern the initial condition, i.e., $MSE_{u_0}$ and $MSE_{v_0}$, were calculated with $N_0 = 256$ coordinates uniformly sampled on the initial pulse and the corresponding real and imaginary parts. Moreover, the PINN network constructed consisted of four hidden layers with 100 neurons, and activated by the tanh function. For solving nonlinear optimization problems like PDEs, the quasi-Newton method is usually more efficient than the gradient descent method.[34] Therefore, the limited Broyden-Fletcher-Goldfarb-Shanno (L-BFGS) optimization algorithm,[35]  which is based on the quasi-Newton method, is applied to accelerate the PINN learning in this work. Specifically, the L-BFGS optimization ends when the $k+1^{th}$ and $k^{th}$ iterations satisfied Equation (8).
\begin{equation}
    \frac{Loss^k-Loss^{k+1}}{\text{max}\{|Loss^k|,|Loss^{k+1}|,1\}}<\epsilon
\end{equation}

\subsection{Multiple Physical Effects in Fiber Modeled by PINN}
\subsubsection{Dispersion}
In a typical optical fiber communication system, $L_D$ and $L_{NL}$ are generally shorter than the fiber length $L$. Thus, the pulse propagation in fiber is mainly affected by dispersion and SPM, which are the two fundamental effects in fiber optics. Before considering the nonlinear effect SPM, it is instructive to first study dispersion effects such as GVD and TOD. For GVD only, the phase of each spectral component is determined by its frequency and propagation distance, and appears as dispersion-induced pulse broadening in the normal dispersion regime ($\beta_2 > 0$), as shown in Figure 3a, which is more evident in the pulse evolution plot (Figure 3b). To well evaluate the model performance, the relative $L_2$ error is adopted as a metric, which is calculated with all reference points ($201\times256$) in the 2D region:
\begin{equation}
    \text{Relative } L_2 \text{ error} = \frac{\sqrt{\sum_{1\leq i\leq 201,1\leq j\leq 256} [h_{pred} (z^i,t^j) - h(z^i,t^j)]^2}} {\sqrt{\sum_{1\leq i\leq 201,1\leq j\leq 256}h^2(z^i,t^j)}}
\end{equation}

As shown in Figure 3c, the relative $L_2$ error is mainly concentrated at the center and edge of the pulse compared with the SSFM results; here, the maximum absolute error was $4.6\times10^{-3}$. When the pulse propagates further, the PINN error remains at an ideal level. For the time- and frequency-domain results shown in Figure 3d and Figure 3e, the maximum error does not exceed $1.4\times10^{-3}$. 
\begin{figure}
	\centering
	\includegraphics[width = 1.0\textwidth]{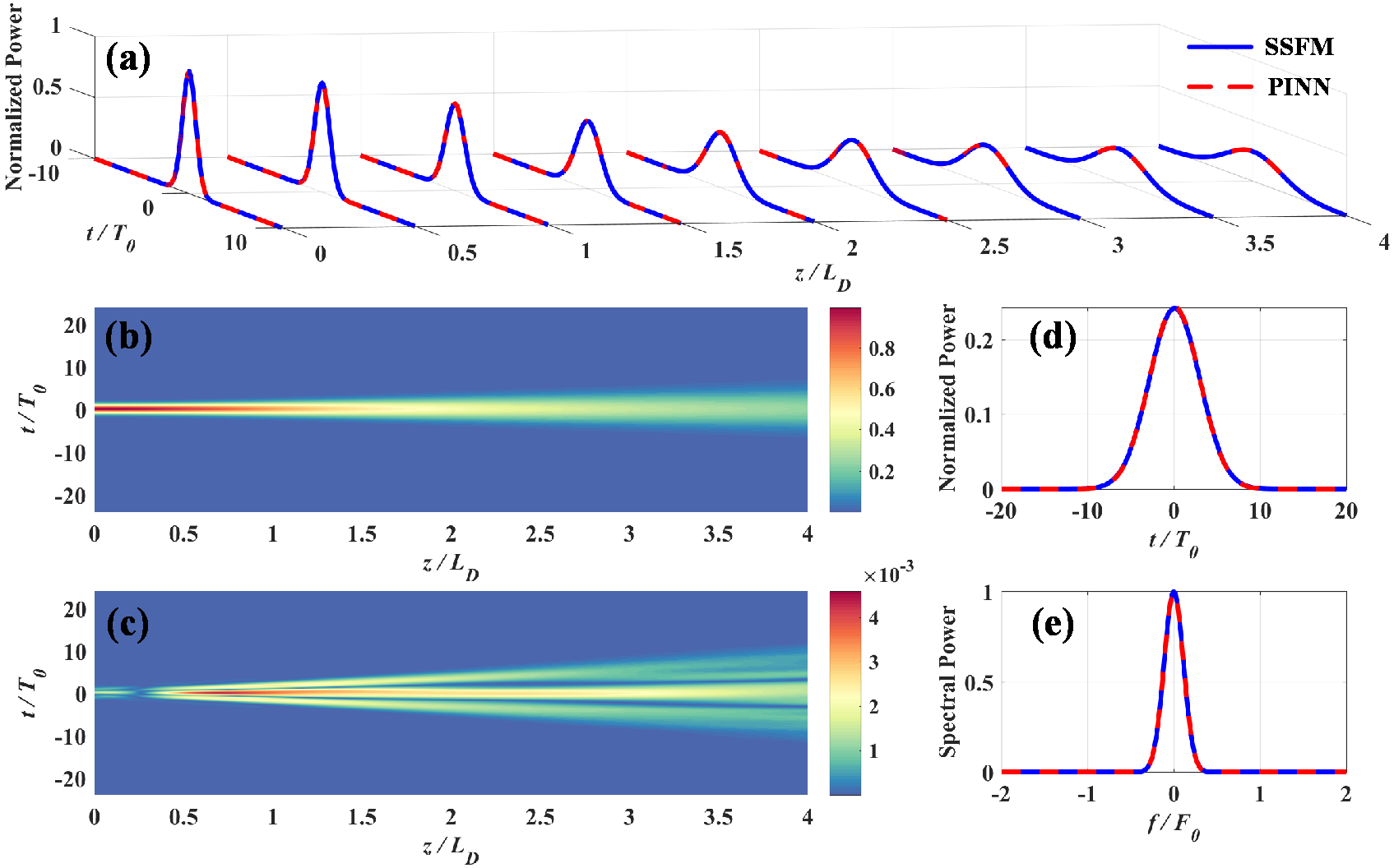}
	\caption{Effects of GVD on pulse propagation: a) waterfall plot for pulse evolution generated by SSFM and PINN, b) pulse evolution generated by PINN, c) error density plot of PINN compared with SSFM, and d) time- and e) frequency-domain results for normalized and spectral powers at $4L_D$, respectively.}
	\setlength{\abovecaptionskip}{-10pt}
    \setlength{\belowcaptionskip}{0pt}

\end{figure}
\raggedbottom

When the pulse wavelength almost coincides with the zero-dispersion wavelength, $\beta_2 = 0$, and TOD must be considered during evolution. As shown in Figure 4a and Figure 4b, oscillations appear near the pulse trailing edge, with the intensity dropping to zero between successive oscillations under the effect of TOD. In fact, these oscillations dampen significantly, even for relatively small $\beta_2$. Similar to the GVD-only case, the error is mainly concentrated at the center and edge of the pulse and always below $5.8 \times 10^{-3}$. Since the dispersion changes the phase of each spectral component only, pulses affected by GVD and TOD differ in the time domain but are consistent with the initial spectrum in the frequency domain, which is apparent from comparison of Figure 4d and Figure 4e with Figure 3d and Figure 3e.
\begin{figure}
	\centering
	\includegraphics[width = 1.0\textwidth]{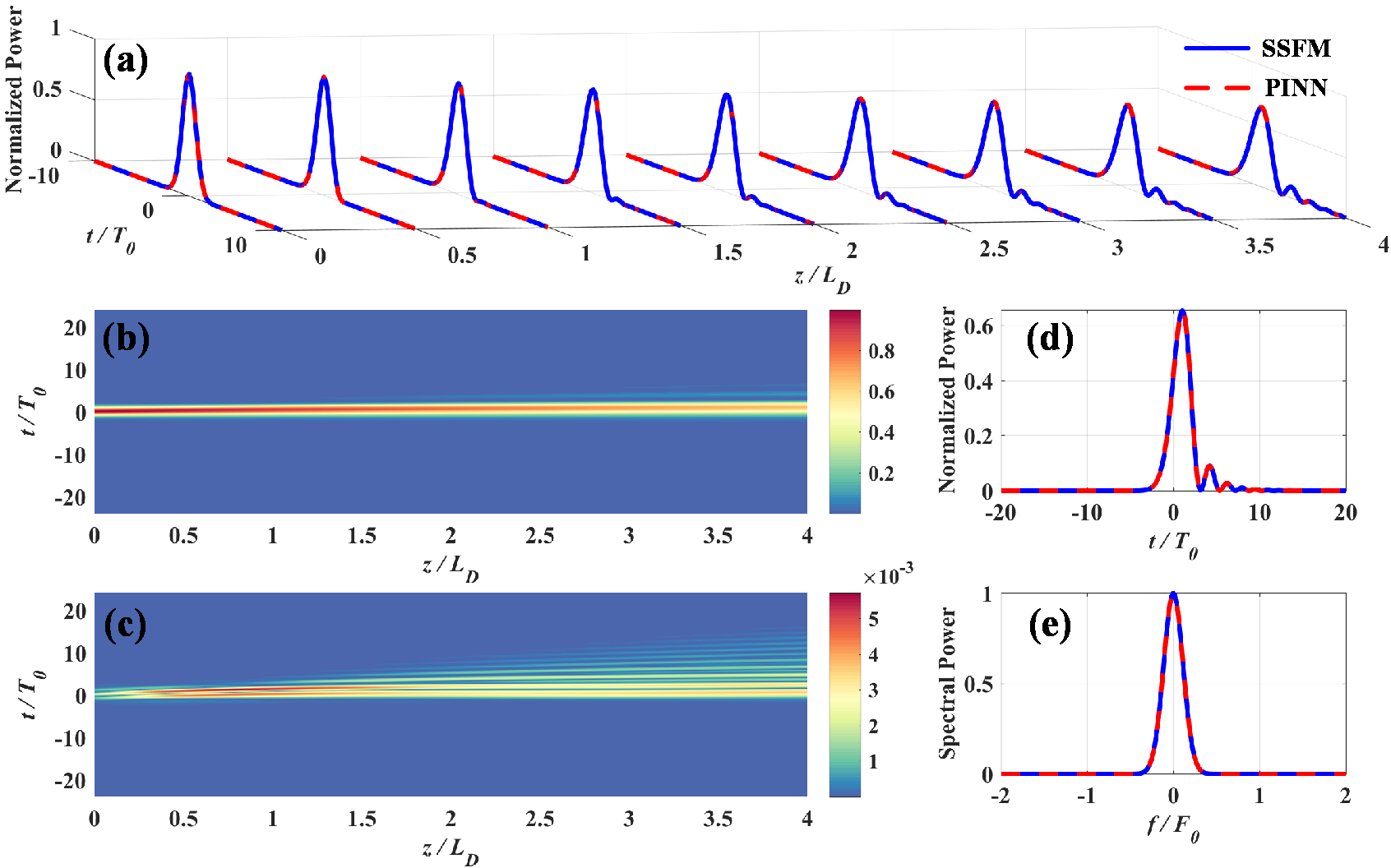}	\setlength{\abovecaptionskip}{-10pt}
    \setlength{\belowcaptionskip}{0pt}
	\caption{Effects of TOD on pulse propagation: a) waterfall plot for pulse evolution generated by SSFM and PINN, b) pulse evolution generated by PINN, c) error density plot of PINN compared with SSFM, and d) time- and e) frequency-domain results for normalized and spectral powers at $4L_D$, respectively.}

\end{figure}
\raggedbottom

\subsubsection{Dispersion and SPM Coexisting}
In general, when a pulse with normal power propagates through a standard fiber, dispersion and SPM coexist in a typical single-mode fiber. Different from the spectrum remains unchanged when only dispersion is considered, the nonlinear effect SPM will increase the complexity on the spectrum evolution. For this more general case, $f$ and $g$ should involve the terms that govern the dispersion and SPM. Figure 5a and Figure 5b show the evolution of the Gaussian pulse in the normal dispersion regime for coexisting GVD and SPM effects. Compared with the evolution in the GVD-only case, the pulse broadening accelerates when SPM is also considered. This behavior can be explained by recognizing that the SPM generates red- and blue-shifted components near the leading and trailing edges of the pulse, and the red-shifted component moves faster in the normal dispersion regime. As shown in Figure 5c, the relative $L_2$ error appears larger initially and diverges during pulse propagation. Although the pulse shape and spectral power are similar to the GVD-only case, as shown in Figure 5d and Figure 5e, the amplitude is significantly lower in the presence of SPM. Note that, for this typical fiber optics scenario, the PINN-based model can effectively model the pulse evolution.
\begin{figure}
	\centering
	\includegraphics[width = 1.0\textwidth]{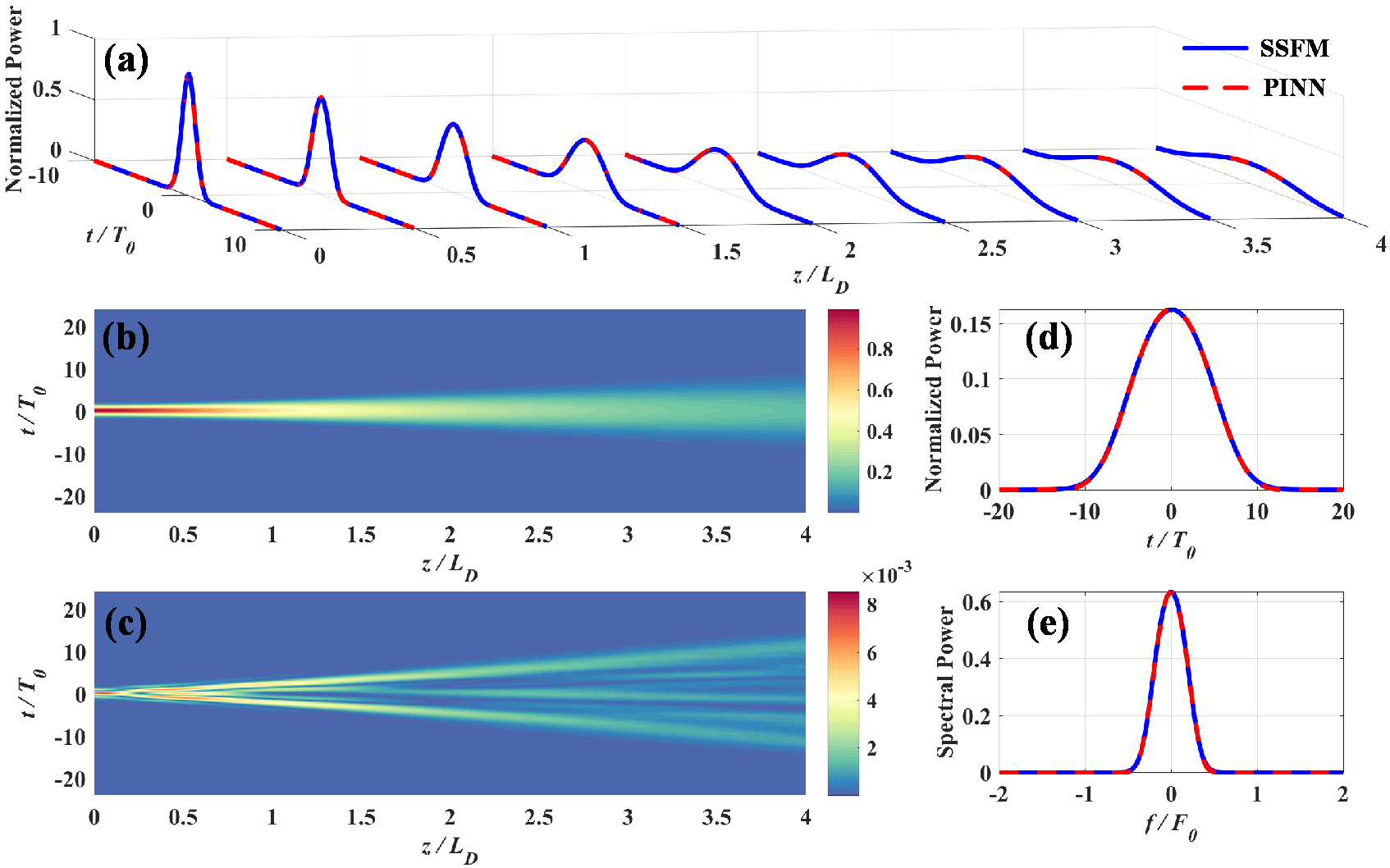}
		\setlength{\abovecaptionskip}{-10pt}
    \setlength{\belowcaptionskip}{0pt}
	\caption{Effects of GVD and SPM on pulse propagation: a) waterfall plot for pulse evolution generated by SSFM and PINN, b) pulse evolution generated by PINN, c) error density plot of PINN compared with SSFM, and d) time- and e) frequency-domain results for normalized and spectral powers at $4L_D$, respectively.}

\end{figure}
\raggedbottom

Unlike the evolution obtained for TOD only, more details appear at the zero-dispersion wavelength case when TOD and SPM coexist. As shown in Figure 6a and Figure 6b, more oscillation peaks appear on the trailing edge of the pulse during evolution. In addition, a non-zero intensity appears at the oscillation minima, as shown in Figure 6d. From the error density plot of Figure 6c, the relative $L_2$ error of the model is slightly higher than the previous cases; here, the maximum error reached $1.56 \times 10^{-2}$. Nevertheless, the evolution under the mixed TOD and SPM action can still be well characterized by PINN. In the optimization process, PINN gradually fitted the main peak, the first oscillation peak, the second oscillation peak, and so on. From the frequency-domain results shown in Figure 6e, the pulse spectrum has an asymmetric double-peak structure, differing considerably from the TOD-only case of Figure 4e, illustrating that TOD introduces spectral asymmetry without affecting the two-peak structure of the SPM. This effect is in sharp contrast to the Figure 5e in which GVD hindered splitting of the spectrum in the normal dispersion regime. 
\begin{figure}
	\centering
	\includegraphics[width = 1.0\textwidth]{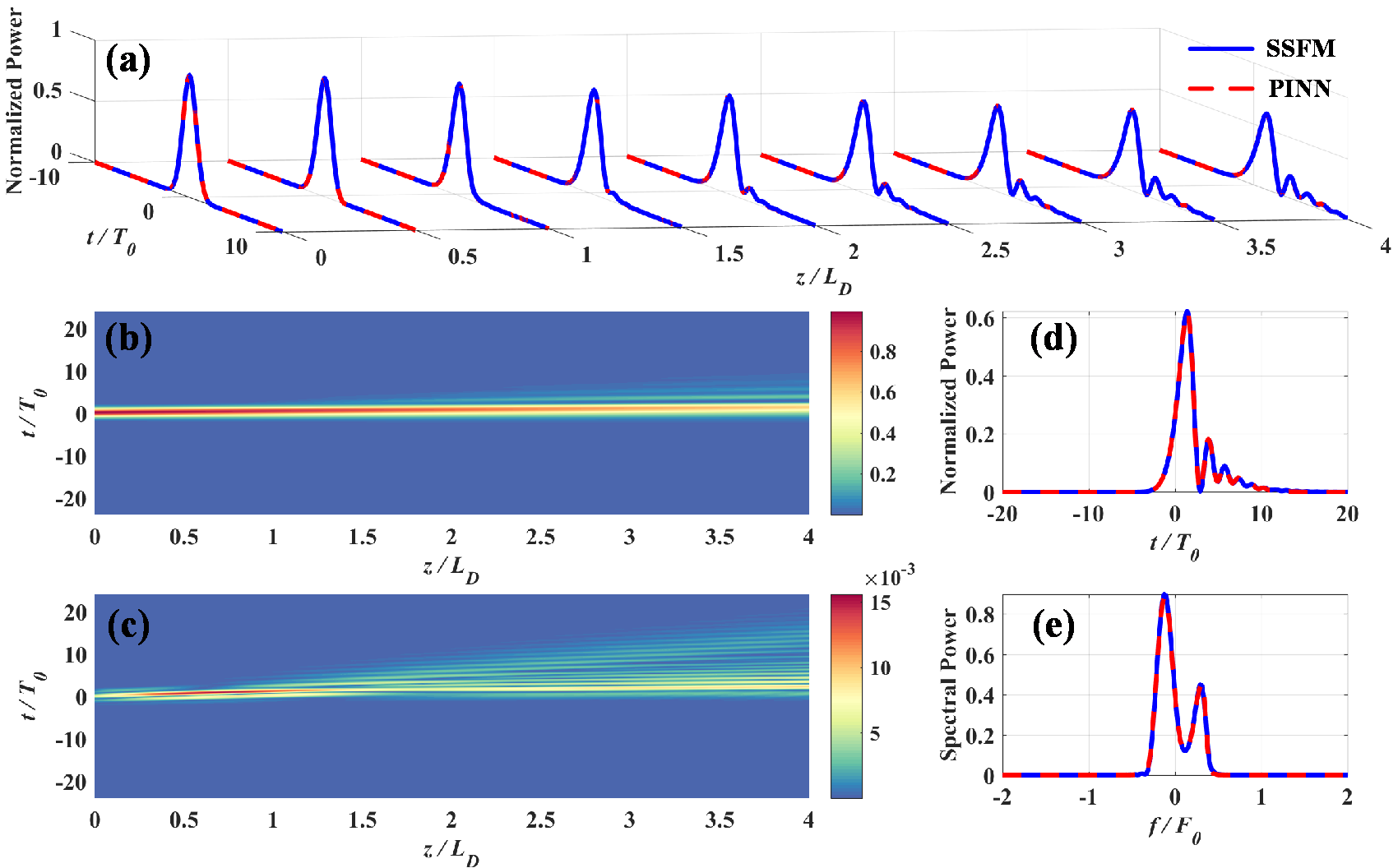}
		\setlength{\abovecaptionskip}{-10pt}
    \setlength{\belowcaptionskip}{0pt}
	\caption{Effects of TOD and SPM on pulse propagation: a) waterfall plot for pulse evolution generated by SSFM and PINN, b) pulse evolution generated by PINN, c) error density plot of PINN compared with SSFM, and d) time- and e) frequency-domain results for normalized and spectral powers at $4L_D$, respectively.}

\end{figure}
\raggedbottom

\subsubsection{Higher-order Nonlinear Effects}
Thus far, we have verified the PINN-based optical fiber modeling in terms of the mixed impact of linear effects and SPM. Next, the higher-order nonlinear effects such as SS and IRS, which are incorporated for ultrashort optical pulses need to be further investigated. Considering the anomalous dispersion case ($\beta_2 < 0$), and keeping the last two terms in the NLSE, the pulse evolutions under SS and IRS effects are shown in Figure 7 and Figure 8, respectively. 

Note that, during modeling of the SS effect, we expanded the distance $z$ to $10L_D$ and set $s$ to $0.2$, to improve the visibility of the pulse shift during evolution. Under the SS effect, the peak moves more slowly than the two sides, as shown in Figure 7a and Figure 7b, and appears shifted toward the trailing side in the anomalous dispersion regime. In particular, the trailing edge becomes steeper with increasing $z$ if anomalous dispersion is neglected. As the propagation progresses, the pulse shape gradually stabilizes, which is more obvious when the initial pulse is a soliton-like pulse $\text{sech}(t)$. As shown in Figure 7(c, d, and e), the error is relatively large and mainly concentrated near the peak in this case, which illustrates the approximation of higher-order differential terms is not easy. In fact, more auxiliary coordinates and exact data can improve model performance, which will be discussed further in Section 4.1.
\begin{figure}
	\centering
	\includegraphics[width = 1.0\textwidth]{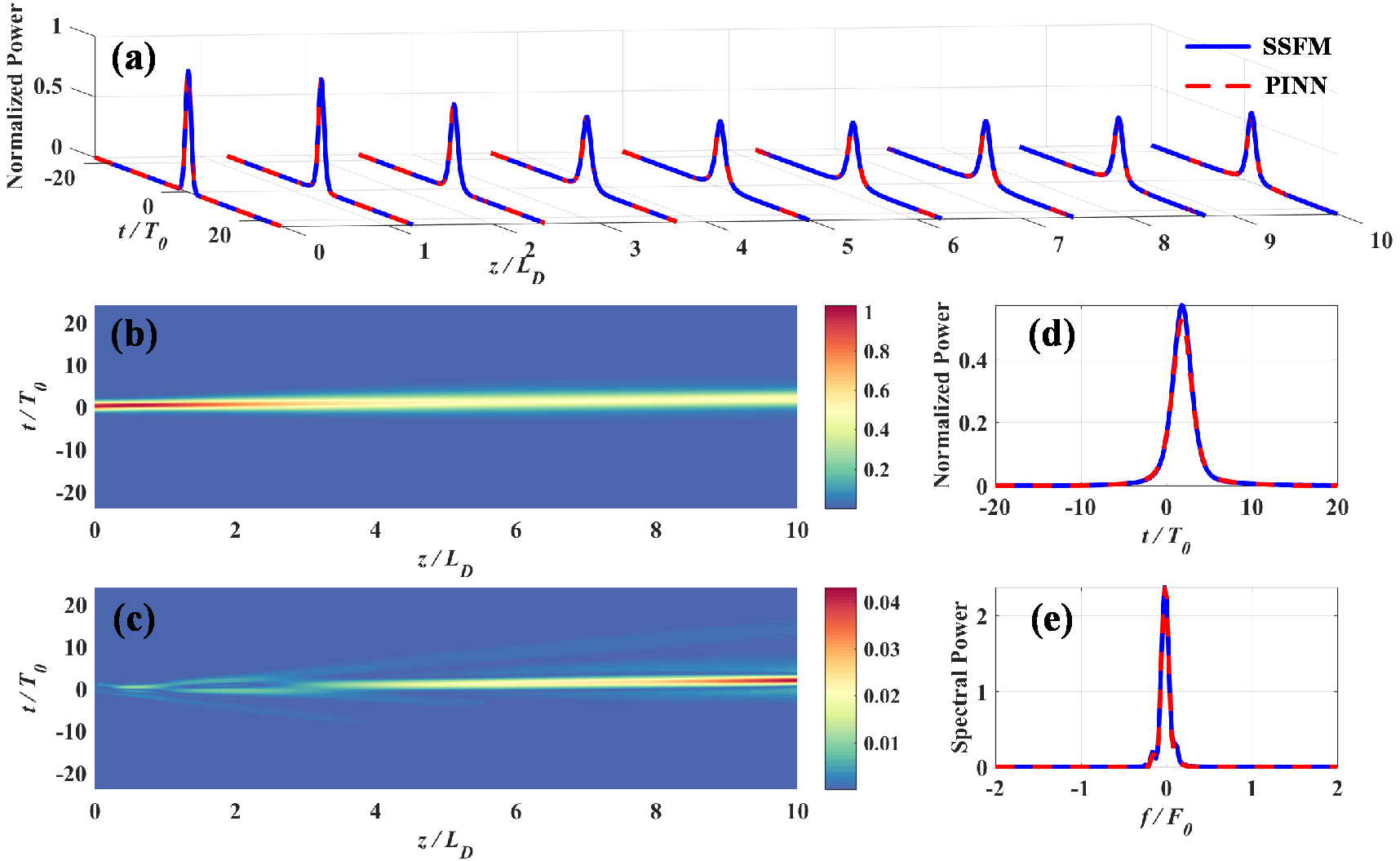}
		\setlength{\abovecaptionskip}{-10pt}
    \setlength{\belowcaptionskip}{0pt}
	\caption{Effects of SS on pulse propagation: a) waterfall plot for pulse evolution generated by SSFM and PINN, b) pulse evolution generated by PINN, c) error density plot of PINN compared with SSFM, and d) time- and e) frequency-domain results for normalized and spectral powers at $4L_D$, respectively.}

\end{figure}
\raggedbottom

For ultrashort optical pulses, IRS becomes quite important. Thus, we set $\tau_R = 0.1$ and $N = 1.4$ to better reflect the effects of IRS on pulse evolution. As shown in Figure 8a and Figure 8b, even if $\tau_R$ is small, it induces pulse fission. As in the previous case, the pulse fission is more obvious when a soliton is transmitted. Moreover, the pulse exhibits a large temporal shift in the time domain and a Raman-induced frequency shift toward longer wavelengths in the frequency domain. These two phenomena are direct consequences of the IRS. From the pulse evolution shown in Figure 8a and Figure 8b, the pulse decelerates after a narrowing process and exhibits a curved trajectory owing to the slower group velocity at longer wavelength.
\begin{figure}
	\centering
	\includegraphics[width = 1.0\textwidth]{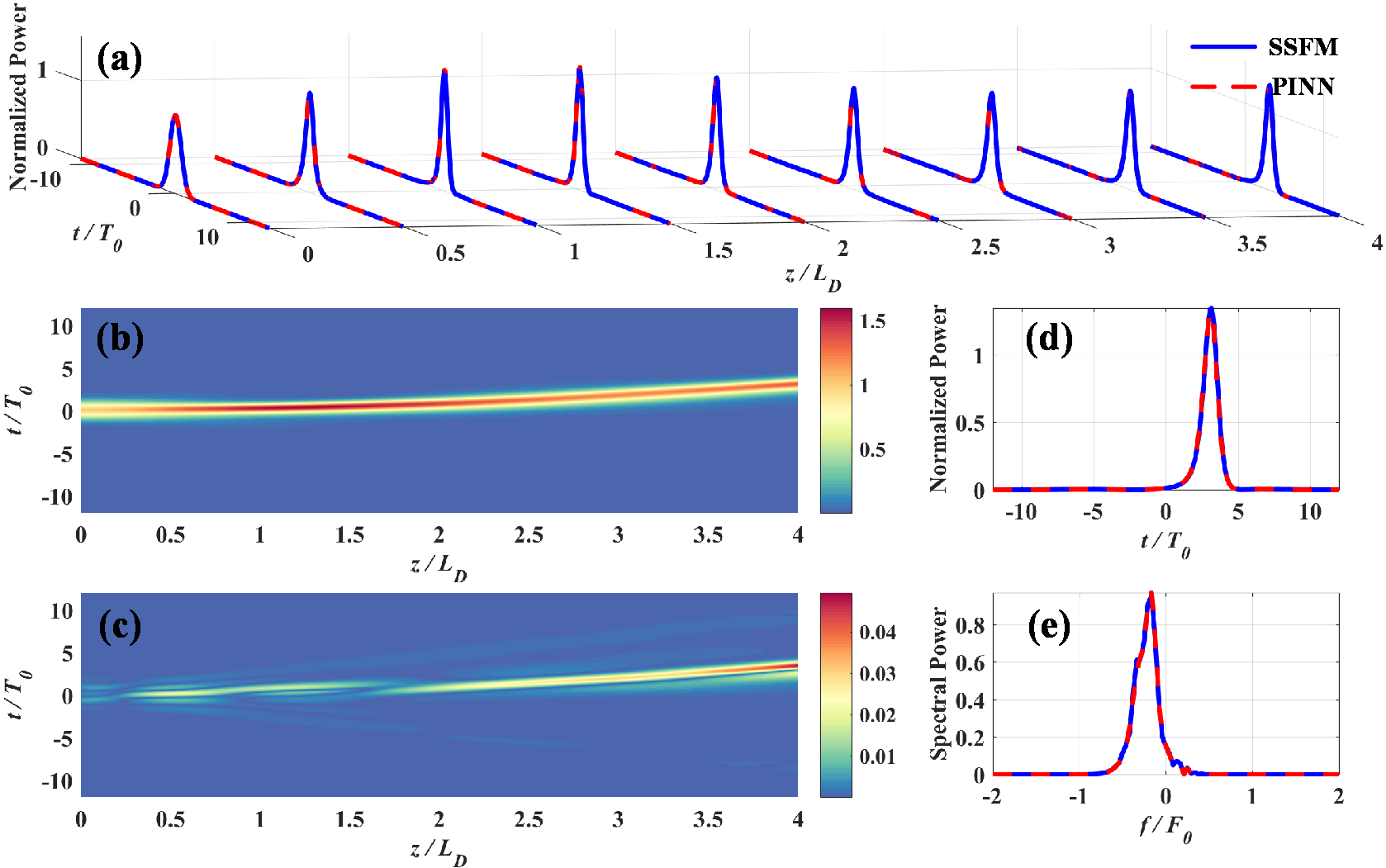}
		\setlength{\abovecaptionskip}{-10pt}
    \setlength{\belowcaptionskip}{0pt}
	\caption{Effects of IRS on pulse propagation: a) waterfall plot for pulse evolution generated by SSFM and PINN, b) pulse evolution generated by PINN, c) error density plot of PINN compared with SSFM, and d) time- and e) frequency-domain results for normalized and spectral powers at $4L_D$, respectively.}

\end{figure}
\raggedbottom

These results indicate that, for the same model configuration, PINN can model these higher-order nonlinear effects effectively. Furthermore, there is only a minimal increase in computational complexity because of the unchanged network structure, which is in contrast to numerical methods such as the SSFM. The relative $L_2$ error of above higher-order nonlinear effects reach the order of $10^{-2}$ with PINN, which can be reduced through further optimization. Despite the error, these effects are well characterized by PINN and do not suffer from the problems caused by the different meshing techniques implemented with numerical methods.

\subsection{PINN-Modeling Universality}
The above verification results reflect the feasibility of PINN-based modeling of nonlinear dynamics in fiber optics based on, but only single-Gaussian initial pulses were considered. However, various types of pulses are employed in practical scenarios for different purposes. Next, to verify the universality of PINN modeling method for different waveforms, herein, some more complex and random pulses typically employed in optical communications were further studied where both the GVD and SPM effects were considered.
\subsubsection{Soliton Propagation}
In the anomalous dispersion regime, the interplay between the dispersion and nonlinear effects creates modulation instability and induces a fascinating pulse-like optical soliton.[36] This type of soliton usually exhibits a periodic evolution pattern. For the fundamental soliton shown in Figure 9, the waveform remains unchanged during the pulse evolution; this feature renders the soliton attractive for ultra-long-distance and ultra-high-capacity optical soliton communication systems. For the setup considered here, the maximum error of the fundamental soliton was $3.9 \times 10^{-3}$, which is even lower than the GVD-only case, as shown in Figure 9c. Clearly, the pulse evolution complexity also affects the model accuracy to a certain extent, apart from the complexity of the governing equations.
\begin{figure}
	\centering
	\includegraphics[width = 1.0\textwidth]{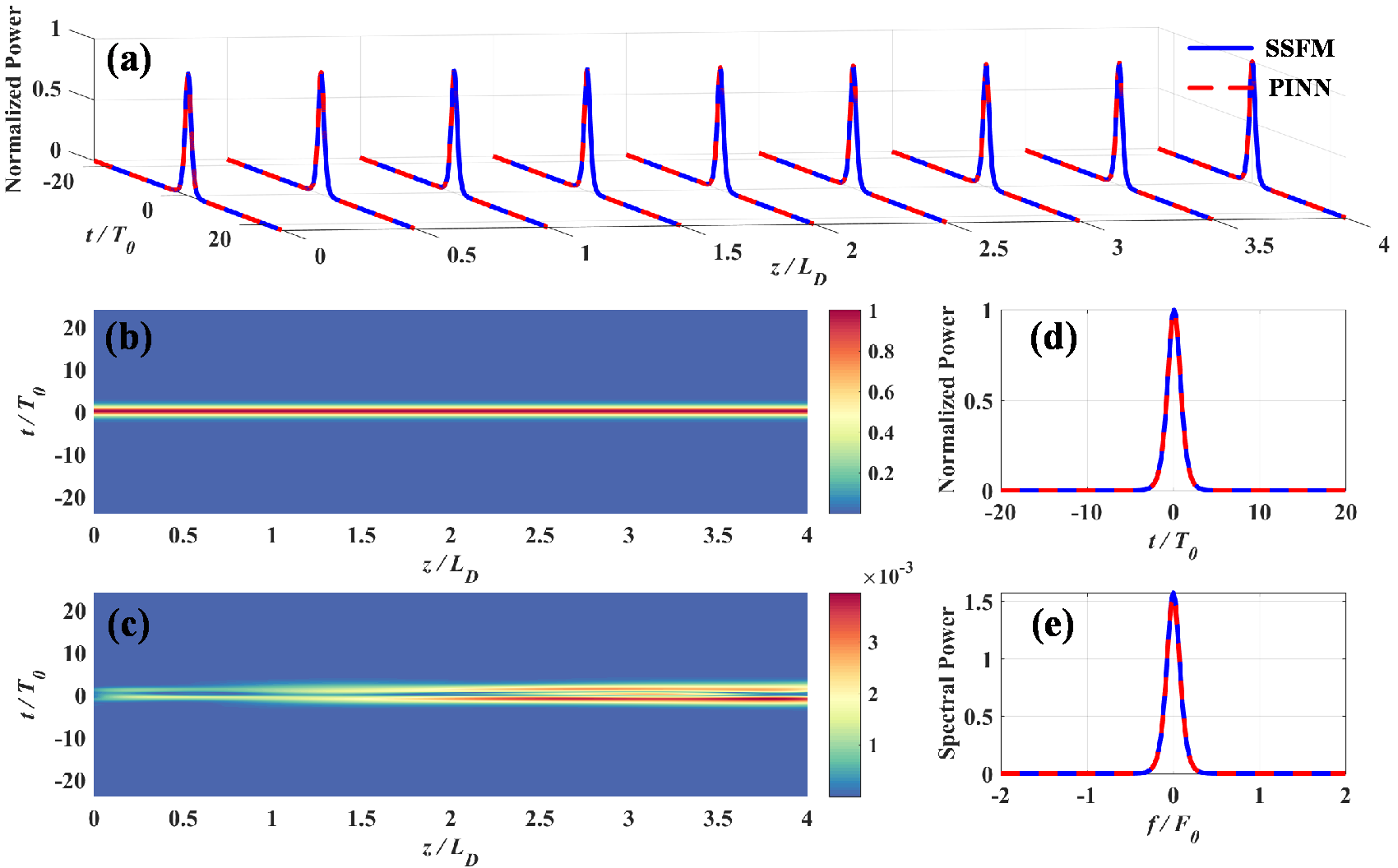}
	\setlength{\abovecaptionskip}{-10pt}
    \setlength{\belowcaptionskip}{0pt}
	\caption{Results for fundamental soliton: a) waterfall plot for pulse evolution generated by SSFM and PINN, b) pulse evolution generated by PINN, c) error density plot of PINN compared with SSFM, and d) time- and e) frequency-domain results for normalized and spectral powers at $4L_D$, respectively.}

\end{figure}
\raggedbottom

The evolution of second-order optical soliton was also studied in this work, as shown in Figure 10. Here, we considered $N = 2$ combined with the initial pulse (i.e., $h(0,t) = 2\text{sech}(t)$), and thus, the governing equations were simplified to the same form as the fundamental soliton. Unlike the fundamental soliton, the second-order soliton presents a periodic evolution pattern with a period of $\pi L_D /2$. As shown in Figure 10c, the error is mainly concentrated in the two peaks because the corresponding pulse power is high, and the influence of pulse power on evolution is further discussed in Section 3.3.1. Except for the slightly lower amplitudes of the pulse centers in Figure 10d and Figure 10e, this periodic case was also well characterized by PINN.
\begin{figure}
	\centering
	\includegraphics[width = 1.0\textwidth]{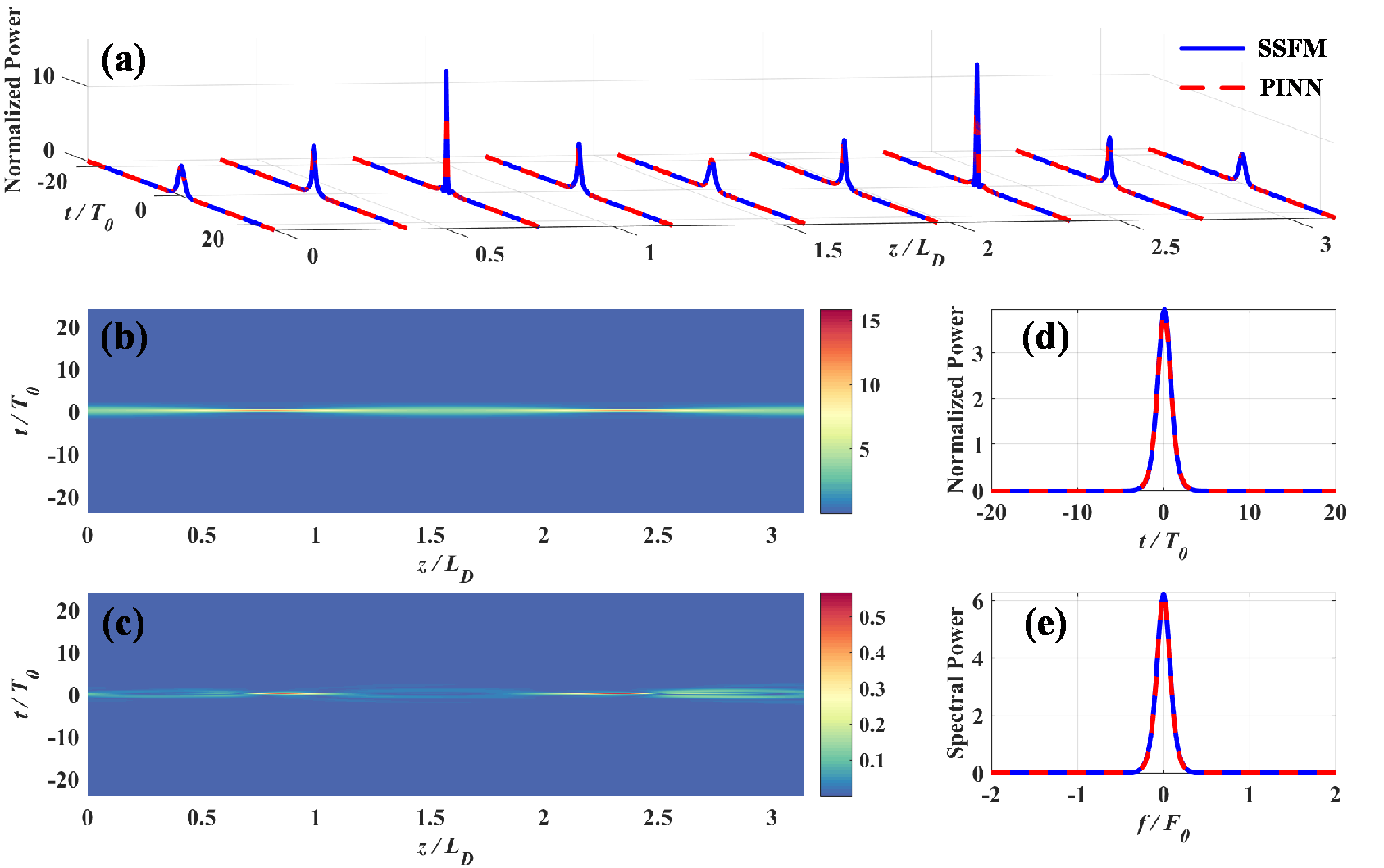}
	\setlength{\abovecaptionskip}{-10pt}
    \setlength{\belowcaptionskip}{0pt}
	\caption{Results for second-order optical soliton: a) waterfall plot for pulse evolution generated by SSFM and PINN, b) pulse evolution generated by PINN, c) error density plot of PINN compared with SSFM, and d) time- and e) frequency-domain results for normalized and spectral powers at $\pi L_D$, respectively.}

\end{figure}
\raggedbottom

\subsubsection{Multi-pulse Propagation }
Apart from above-studied single-pulse cases, another essential case in practical application is multi-pulse propagation in fiber. Here, two complex waveforms composed of 5 single-Gaussian pulses were investigated: regular waveform with equal amplitudes (Figure 11) and random with non-equal amplitudes (Figure 12). These two waveforms can be viewed as two common signals in pulse-amplitude modulation (PAM). 
For the equal-amplitude waveform in Figure 11, all the single pulses suffer from the broadening and crosstalk occurs among each other as the propagation progresses, resulting in the newborn peaks between adjacent pulses. For this type of transmission system with inter-pulse interference, the PINN modeling error reached the level of $10^{-2}$ for the setup considered here, as shown in Figure 11c. The error is relatively large at the peaks, especially the central peak, as shown in Figure 11d. This is because the broadening of each pulse is superimposed at these peaks, and the central region is most affected when all pulses have equal amplitudes.
\begin{figure}
	\centering
	\includegraphics[width = 1.0\textwidth]{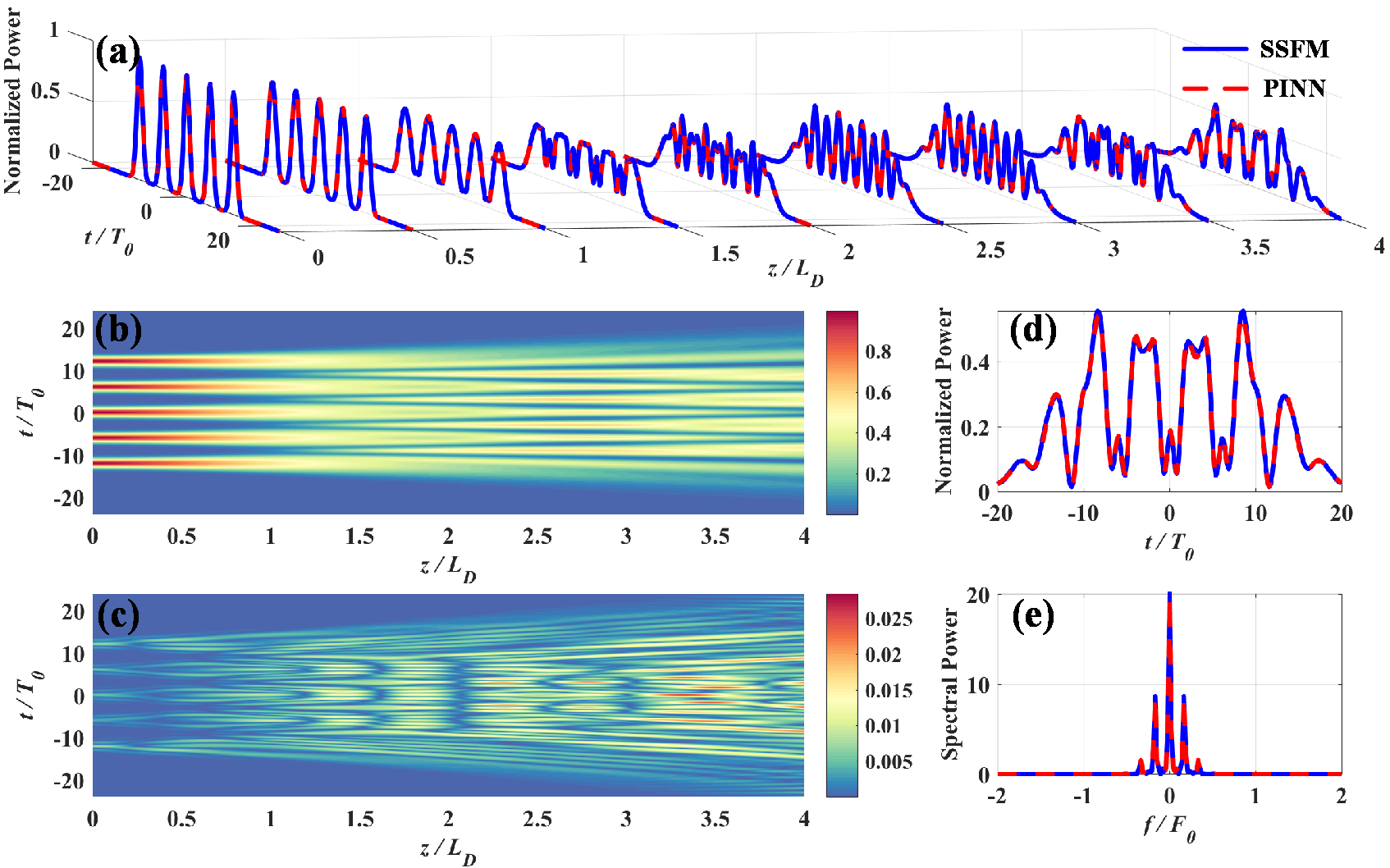}
	\setlength{\abovecaptionskip}{-10pt}
    \setlength{\belowcaptionskip}{0pt}
	\caption{Results for complex waveform of 5 single-Gaussian pulses with equal amplitudes:  a) waterfall plot for pulse evolution generated by SSFM and PINN, b) pulse evolution generated by PINN, c) error density plot of PINN compared with SSFM, and d) time- and e) frequency-domain results for normalized and spectral powers at $4L_D$, respectively.}

\end{figure}
\raggedbottom

Inter-pulse crosstalk also appears in the non-equal-amplitude waveform shown in Figure 12. However, the pulse evolution is no longer symmetrical, and the degree of inter-pulse crosstalk varies with the pulse amplitude. Because the initial pulses near the leading edge have lower amplitudes, the error of the peaks near the leading edge is relatively large during evolution, as indicated by the error density plot shown in Figure 12d. The maximum error of the time-domain result was $2.16 \times 10-2$, which is slightly higher than that of the single-pulse case. For this case where the pulse evolution details are richer, the PINN model can still achieve such good results with the same configuration, and the model performance can be further improved when more coordinate points and exact data are available. From this preliminary attempt on multi-pulse investigation, the satisfactory results indicate the considerable potential of this approach for arbitrary waveforms modeling. 
\begin{figure}
	\centering
	\includegraphics[width = 1.0\textwidth]{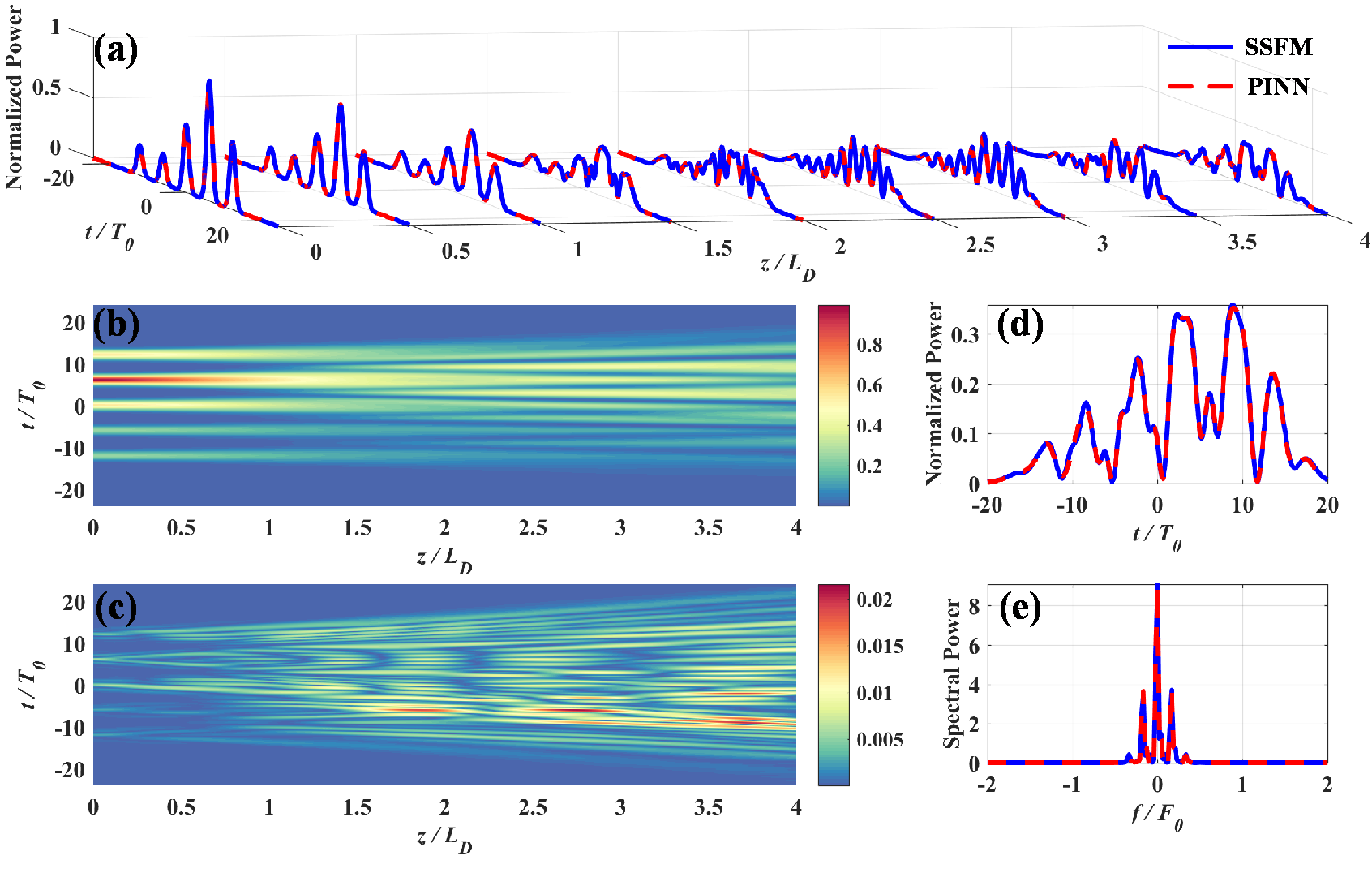}
	\setlength{\abovecaptionskip}{-10pt}
    \setlength{\belowcaptionskip}{0pt}
	\caption{Results for complex waveform of 5 single-Gaussian pulses with non-equal amplitudes:   a) waterfall plot for pulse evolution generated by SSFM and PINN, b) pulse evolution generated by PINN, c) error density plot of PINN compared with SSFM, and d) time- and e) frequency-domain results for normalized and spectral powers at $4L_D$, respectively.}

\end{figure}
\raggedbottom

\subsection{The Generalizability of PINN Modeling}
At present, most of PINN-based PDE solutions are still hindered by insufficient generalizability, that is, a constructed PINN model is only feasible and suitable to one scenario. For different scenarios, the physical constraints change accordingly. Thus, a different optical fiber model must be developed with a new loss function. In the previous studies, the loss function of PINN uniquely corresponds to the physical constraints of one scenario in the modeling process, resulting in the poor generalizability. However, if multiple physical constraints corresponding to multiple scenarios could be introduced to the loss function during the training process, a multi-scenario PINN model could be realized without remodeling. Here, to verify the feasibility of generalizable PINN model for multi-scenario modeling, the optical fiber model was trained under the effects of GVD and SPM using pulses of different powers and random waveforms composed of three sub-pulses, as shown in Figure 13.
\begin{figure}
	\centering
	\includegraphics[width = 1.0\textwidth]{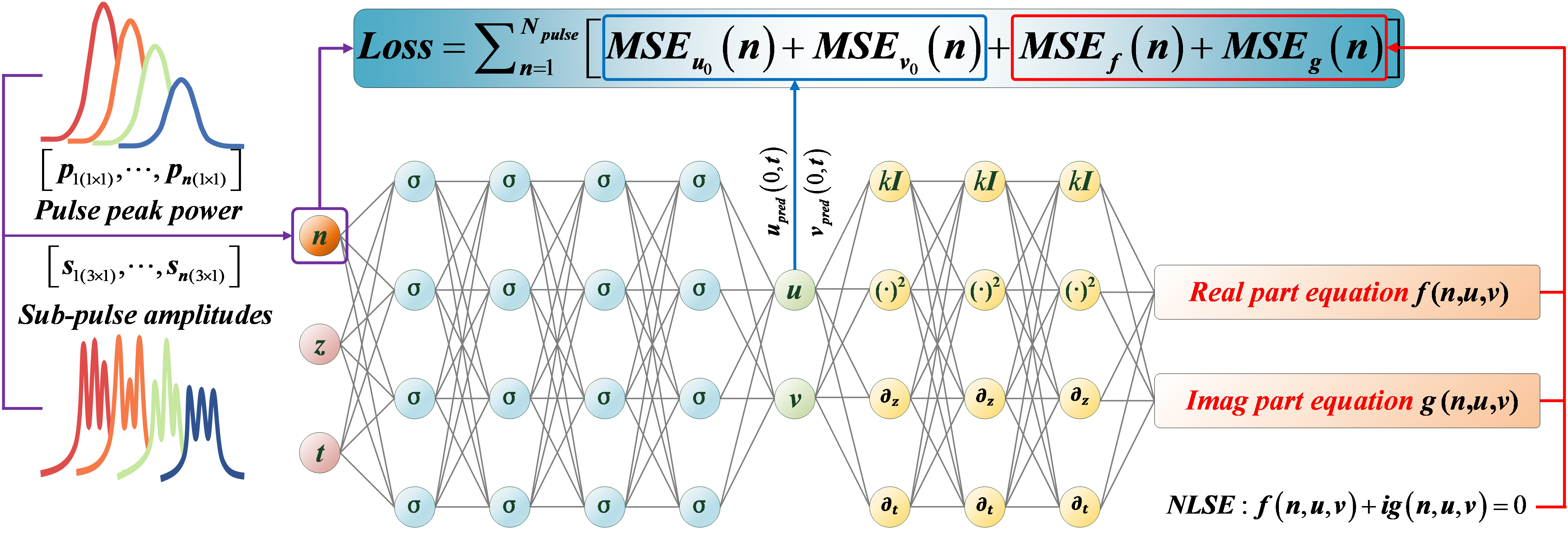}
	\setlength{\abovecaptionskip}{-10pt}
    \setlength{\belowcaptionskip}{0pt}
	\caption{Generalizability modeling of PINN for scenarios with multi-power / random waveforms in optical fibers.}

\end{figure}
\raggedbottom

\subsubsection{Modeling Generalizability for Multi-power Scenario in Optical Fiber}
As detailed in Figure 13, an NLSE-independent variable $p$ representing the initial-pulse peak power is introduced to the PINN-network input layer to achieve model generalizability. This introduced power variable can be regarded as a physical controller mapping different cases to their own loss function. Specifically, each auxiliary coordinate $(z^i,t^i)$ is combined with the peak power $p^i$. Hence, a new input vector $(z^i,t^i,p^i)$ is formed, with the output being the real and imaginary parts of the complex envelope at $(z^i,t^i)$ with $p^i$. Each case uniquely corresponds to its specific initial condition, and the new loss function can be expressed in the form of the sum of multiple loss functions, as illustrated in Figure 13. 

Next, taking $p = 1$ as a reference, 11 Gaussian pulses in the set $[0.5, 1.5]$ with a step size $\Delta p = 0.1$ were selected as initial pulses. The loss function considered the physical constraints of 6 pulses ($p = [0.5, 0.7, 0.9, 1.1, 1.3, 1.5]$), and the remaining 5 pulses were used to verify the power generalizability of PINN. From Figure 14a, the relative $L_2$ error in the constructed temporal-spatial domain was less than $1.8 \times 10^{-2}$ for this setup, regardless of whether physical constraints were considered in the loss function. In this case, the error increases with the peak power, consistent with the second-order optical soliton case discussed in Section 3.2.1. In fact, this phenomenon is normal, as the PINN is essentially a neural network based on a multilayer perceptron and its performance degrades for higher data value, which reflects the importance of data normalization. Moreover, the errors of the 5 verification pulses are all located between adjacent points, indicating the strong generalizability of the PINN model. Figure 14b and Figure 14c show the error density plots for $p = 1.4$ and $p = 1.5$, corresponding to cases where the initial pulse constraints are excluded from and included in the loss function, respectively. The error distributions are almost the same, except for the slight difference in amplitude (the error for $p = 1.5$ is greater). This result confirms that the PINN model generalizes the pulse evolution in cases with different powers.
\begin{figure}
	\centering
	\includegraphics[width = 1.0\textwidth]{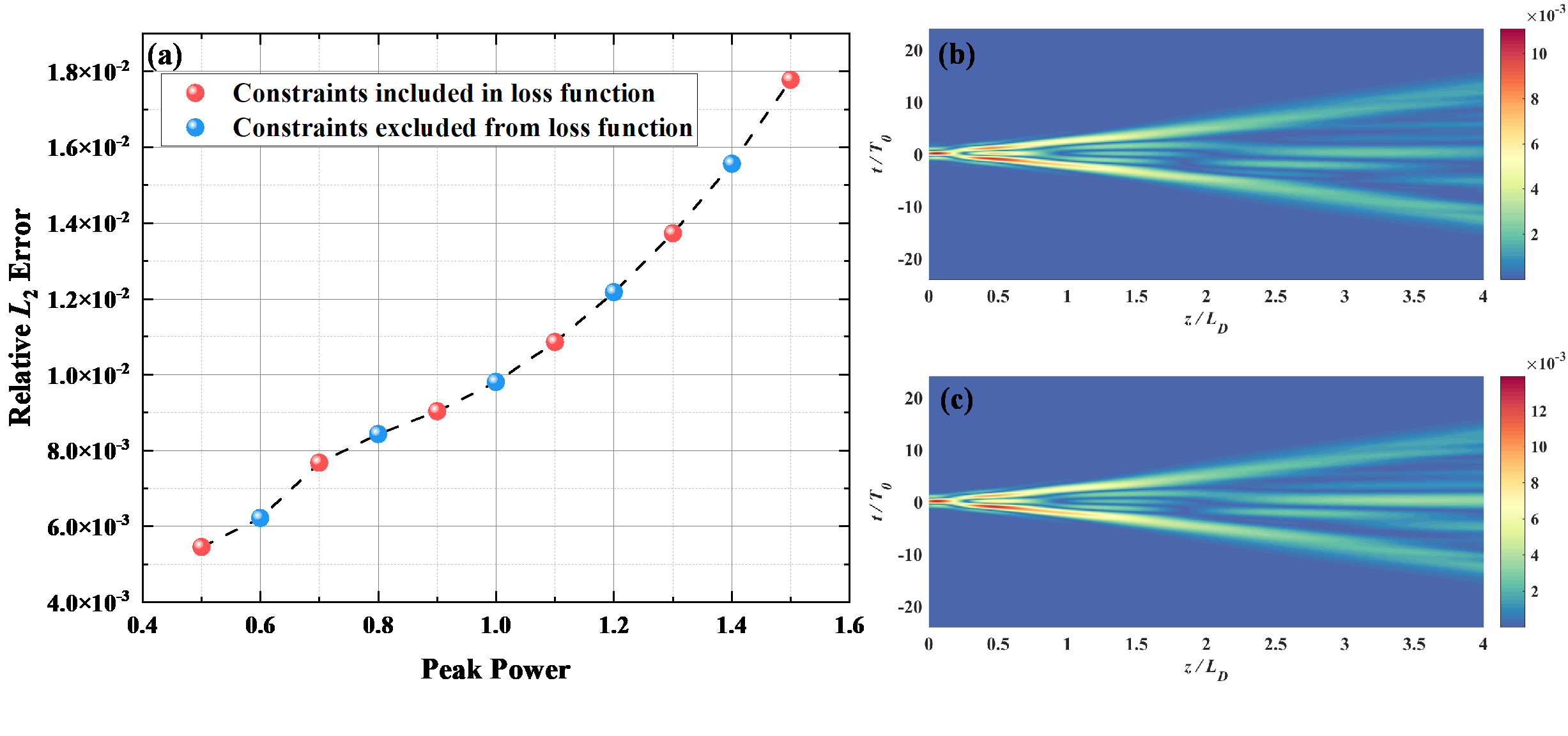}
	\setlength{\abovecaptionskip}{-10pt}
    \setlength{\belowcaptionskip}{0pt}
	\caption{Generalizability demonstration: a) relative $L_2$ error according to peak power with constraints included in / excluded from the loss function and error density plots with b) $p = 1.4$ and c) $p = 1.5$, respectively.}

\end{figure}
\raggedbottom

\subsubsection{Modeling Generalizability for Random Waveforms in Optical Fiber}
The above results and discussions illustrate the PINN generalizability in multi-power scenarios. However, random waveform transmission is more general in practice, and even more complicated. Further, we considered transmitted waveforms composed of 3 Gaussian pulses with amplitudes randomly selected from the set $[0.25, 0.5, 0.75, 1]$. Note that these waveforms can be regarded as PAM4-format signals transmitted in an optical fiber communication system. As above, the amplitudes of the 3 sub-pulses were introduced as the physical parameter controllers to the input layer, allowing PINN to learn the evolution of the corresponding waveform via the process shown in Figure 13. Among all possible 34 waveforms, 10 random waveforms were selected as the sampled scenarios for PINN to learn. Note that all 4 amplitudes should be included in the selected 10 waveforms to increase the model performance. The loss function was still in the sum of multiple loss functions, including the initial constraints of the 10 selected waveforms.

\begin{figure}
	\centering
	\includegraphics[width = 1.0\textwidth]{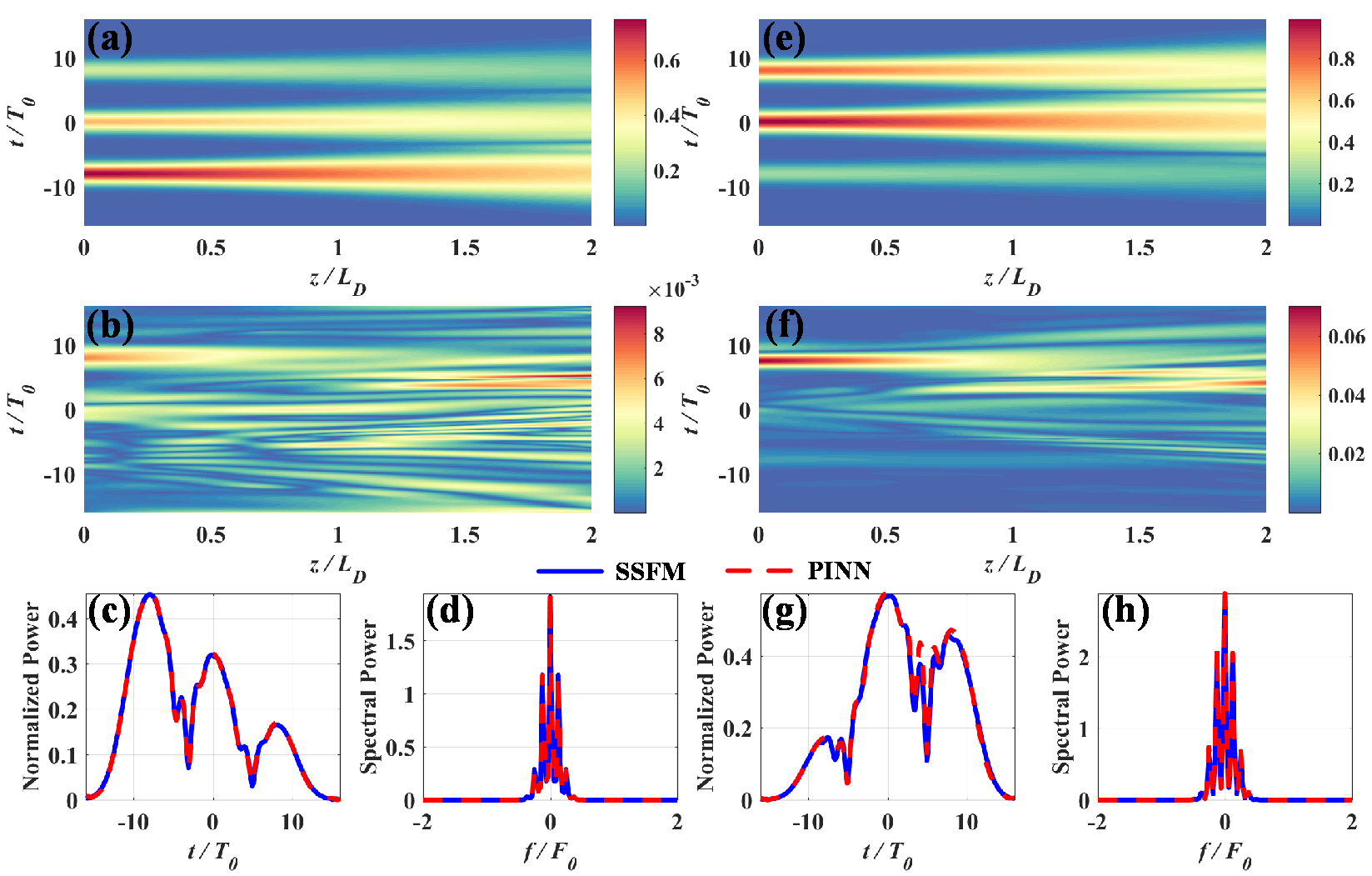}
		\setlength{\abovecaptionskip}{-10pt}
    \setlength{\belowcaptionskip}{0pt}
	\caption{Modeling generalizability for random waveforms in optical fiber: a) pulse evolution generated by PINN, b) error density plot of PINN compared with SSFM and c) time- and d) frequency-domain results for normalized and spectral powers at $2L_D$ for random waveform composed of 3 sub-pulses with amplitudes $[0.75, 0.5, 0.25]$. e) Pulse evolution generated by PINN, f) error density plot of PINN compared with SSFM and g) time- and h) frequency-domain results for normalized and spectral powers at $2L_D$ for random waveform composed of 3 sub-pulses with amplitudes $[0.25, 1, 0.75]$.}

\end{figure}
\raggedbottom

As shown in Figure 15, the PINN-based model effectively characterized the evolution of the two unlearned waveforms. Here, the random waveform composed of 3 sub-pulses with amplitudes $[0.75, 0.5, 0.25]$ was similar to a waveform (composed of 3 sub-pulses with amplitudes $[0.75, 0.75, 0.25]$) learned by the PINN, indicating a relatively small error. In contrast, the random waveform composed of 3 sub-pulses with amplitudes $[0.25, 1, 0.75]$ differs considerably from the 10 learned waveforms, indicates a relatively large error. Nevertheless, the overall error could be controlled to the order of $10^{-2}$. The results in both the time and frequency domains are almost consistent with those generated by the SSFM (Figure 15 (c, d, g, and h)), even though the PINN network never learned those scenarios before. This outcome presents the robust generalizability of PINN and illustrates that the law of random signals in PAM4 format can be effectively learned by PINN. From this perspective, the main significance of PINN is its potential equivalence to a complex operator for solution of complex PDE with arbitrary physical constraints. The PINN-based model established for the above problem can be regarded as an NLSE operator that predicts the evolution of random signals in PAM4 format. Furthermore, PINN has the potential to act as an NLSE operator to characterize random signal evolution in any modulation format.

\section{Analysis and Discussions}
\subsection{PINN-modeling Accuracy Analysis}
A major feature of PINN is to calculate the governing equations through auxiliary coordinates, so that the model meets the constraints of PDE. Therefore, the number of auxiliary coordinates has a great influence on the model accuracy. Another factor that affects the accuracy is the strength of the physical constraints, and the model performance will be further improved when physical constraints are stronger. In the previous results, only initial conditions were considered. Figure 16 shows the model performance changing with the auxiliary coordinates number under several different physical constraints when GVD and SPM coexist. It can be seen that the model performance increases with the number of auxiliary coordinates and 5000 auxiliary coordinates are enough to reach a good accuracy for our problem. Moreover, the impact of physical constraints on the model performance is also intuitive, as shown in Figure 16. The model error will be one order of magnitude higher than the other two cases when only initial pulse is considered. When considering pulses at three distances simultaneously, the optimal model performance is achieved. In fact, when the physical constraints are stronger, the model performance will be further improved, and these constraints can be the exact data or prior knowledge such as symmetry conditions.
\begin{figure}
	\centering
	\includegraphics[width = 0.5\textwidth]{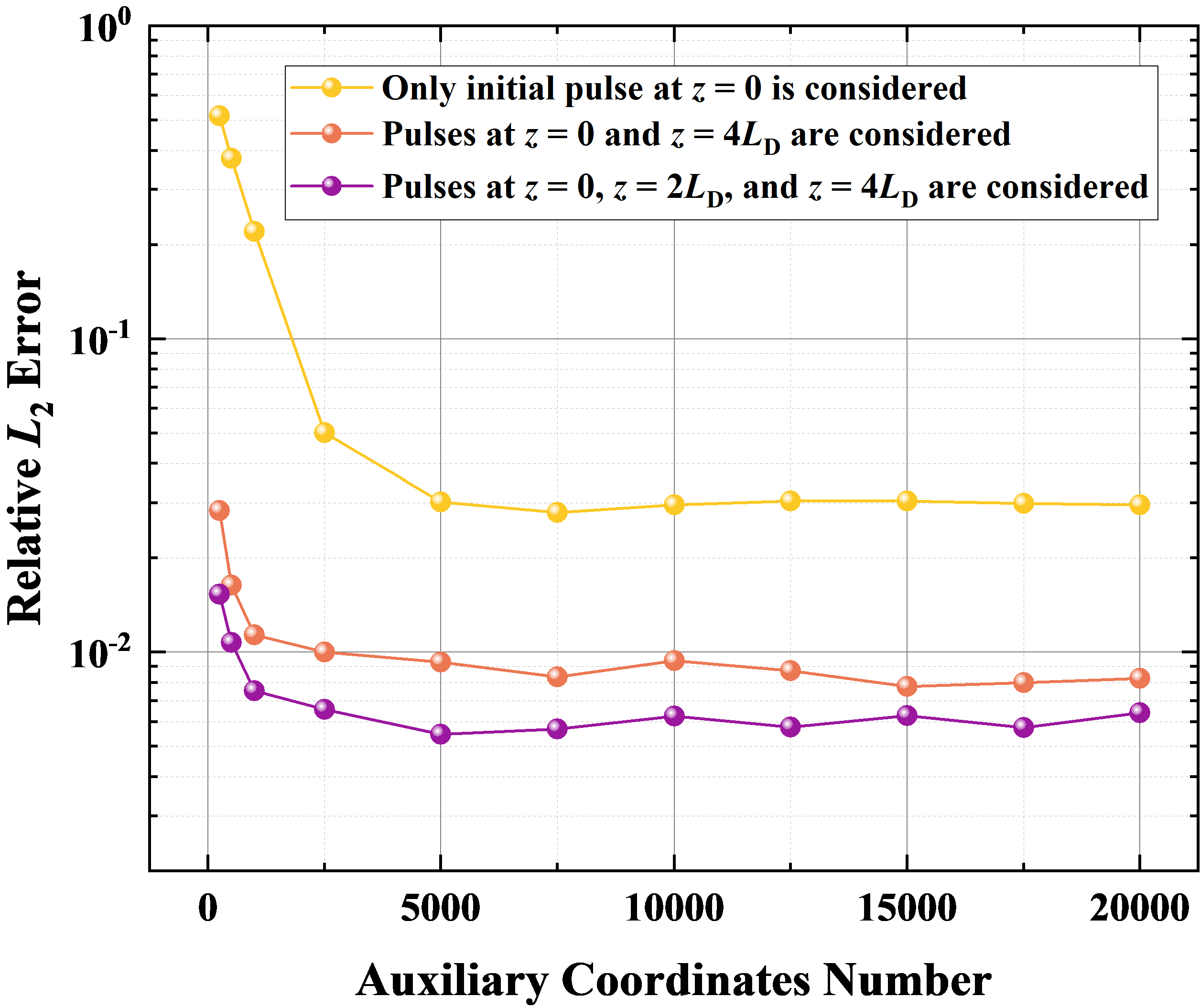}
		\setlength{\abovecaptionskip}{0pt}
    \setlength{\belowcaptionskip}{0pt}
	\caption{Relative$L_2$ error of the PINN-based model with the auxiliary coordinates number under several different physical constraints when GVD and SPM coexist.}

\end{figure}
\raggedbottom

  Next, we compared the model performance of PINN with the data-driven neural network (DDNN). The DDNN for comparison consisting of 4 hidden layers with 100 neurons was the same as PINN. Tanh and L-BFGS were also adopted for activation and optimization. Similarly, DDNN took $(z,t)$ as input and $(u,v)$ as output. In the constructed region, the exact data at equal distance was selected for data-driven modeling and as physical constraints of PINN. After the optimization of L-BFGS, DDNN fully fitted the exact data, while PINN also learned the governing equations with auxiliary coordinates at the same time. To avoid accidents, we recorded 5 sets of results with two methods, and calculated the average value as the final results. Figure 17 shows the performance comparison between DDNN and PINN, where 20,000 auxiliary coordinates were used for PINN modeling. When the data volume is quite small, the error of DDNN is relatively large, and impossible to model the pulse evolution. Specifically, DDNN requires the exact data of at least 5 pulses to characterize the pulse evolution well. Furthermore, the performance of PINN is improved with the increase of data, which illustrates that more exact data contributes to the better performance of PINN. Note that prior knowledge such as the governing equations is still the decisive factor affecting the model performance, which can be seen from the comparison between PINN and DDNN: even with large volume of data, the performance of DDNN was still worse than PINN with only initial and received pulses as constraints.
\begin{figure}
	\centering
	\includegraphics[width = 0.5\textwidth]{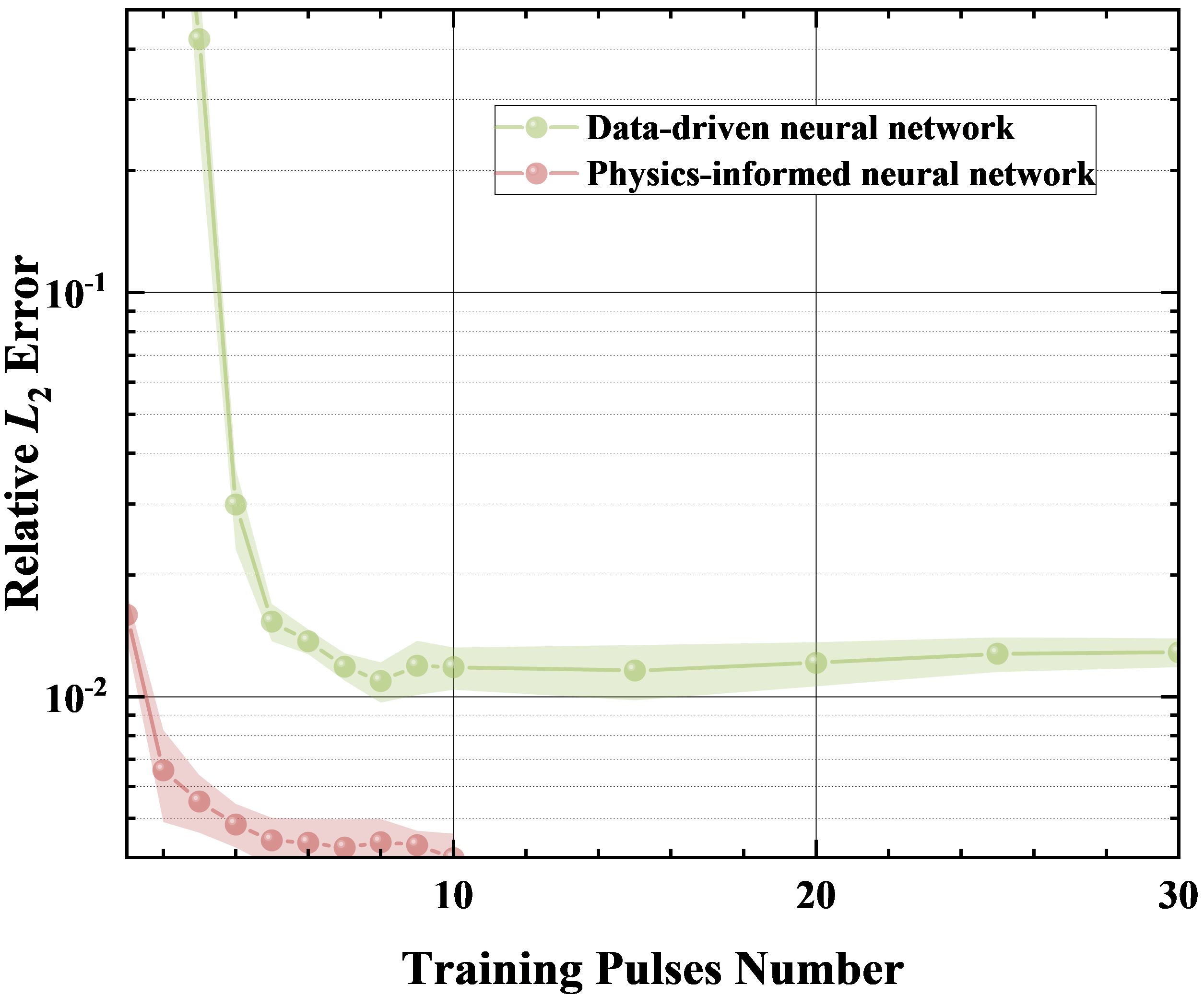}
		\setlength{\abovecaptionskip}{0pt}
    \setlength{\belowcaptionskip}{0pt}
	\caption{Relative$L_2$ error of PINN vs the data-driven neural network as a function of the number of training pulses.}

\end{figure}
\raggedbottom

\subsection{PINN-modeling Complexity Analysis}
Finally, the comparative analysis on the computational complexity of PINN and SSFM was conducted, where a theoretical calculation of the multiplication operations in the modeling process is required. Such calculations are commonly applied to hardware complexity comparisons among different algorithms.[37] For SSFM, the calculation performed in each step includes two fast Fourier transform (FFT) operations, and each FFT operation contains $2N_{FFT}log_2N_{FFT}$ real-valued multiplications,[38,39] where $N_{FFT} = N_0$ is the FFT size. Considering the entire pulse evolution process in the fiber and ignoring the exponential operation, the multiplications of SSFM can be expressed as:
\begin{equation}
    C_{SSFM}=N_{step}(4N_{FFT}log_2N_{FFT}+k_1N_{FFT}+k_2N_{FFT}^2)
\end{equation}
where $N_{step}$ denotes the number of steps into which the entire fiber is divided. This value must be properly set according to the propagation distance and accuracy requirements. Further, $k_1$ and $k_2$ denote the number of rest multiplications proportional to $N_{FFT}$ and $N_{FFT}^2$ in each step, respectively. From Equation (10), the SSFM complexity increases with the FFT size and increases linearly with $N_{step}$.

The multiplication operation is also the main source of the PINN computational complexity. If the $i^{th}$ layer of the PINN network contains $n_i$ neurons, $(n_1n_2 + n_2n_3 +\cdots+ n_{M-1}n_M)$ multiplications are required for a fully connected neural network with $M$ layers. In addition, all the terms in loss function must be calculated for each coordinate. If the number of multiplication operations in this part is $k_3$, and the multiplications of PINN can be expressed as:
\begin{equation}
    C_{PINN}=(N_0+N_f)(n_1 n_2 + n_2 n_3 + \cdots+ n_{M-1} n_M+ k_3)
\end{equation}

It is apparent from Equation (11) that the PINN complexity increases linearly with the number of selected coordinates, which is quite similar to the linear increase in SSFM with increasing steps. To facilitate a comparison of the computational complexity of the two methods, calculations were performed for the same data volumes (i.e., $N_0 + N_f = N_{FFT} \times N_{step}$). For the two models, the relationship between the number of multiplications and the propagation distance was plotted considering all parameters, as shown in Figure 18. An appropriate step size was set to ensure the accuracy of the two methods. Since the data volume was proportional to the propagation distance, the number of multiplications increased linearly with the propagation distance. Moreover, $N_0$ had far greater influence on the SSFM complexity than that of PINN, as the former involved a large number of multiplication operations with $O(n^2)$ complexity in the nonlinear calculations. When $N_0=512$, the multiplications of SSFM and PINN were approximately $2.14 \times 10^{10}$ and $3.11 \times 10^9$ for a distance of $10L_D$ (approximately 1000 km in a standard single-mode fiber). Thus, the computational complexity of the PINN-based model was only approximately 14.6\% that of the SSFM-based model. In fact, the data volume actually required by PINN is not as much as SSFM, so the computational complexity of PINN can be even lower. Furthermore, the number of PINN multiplications is mainly affected by the neural network structure rather than the data volume, and thus, PINN is more efficient in the case with large data volume.
\begin{figure}
	\centering
	\includegraphics[width = 0.5\textwidth]{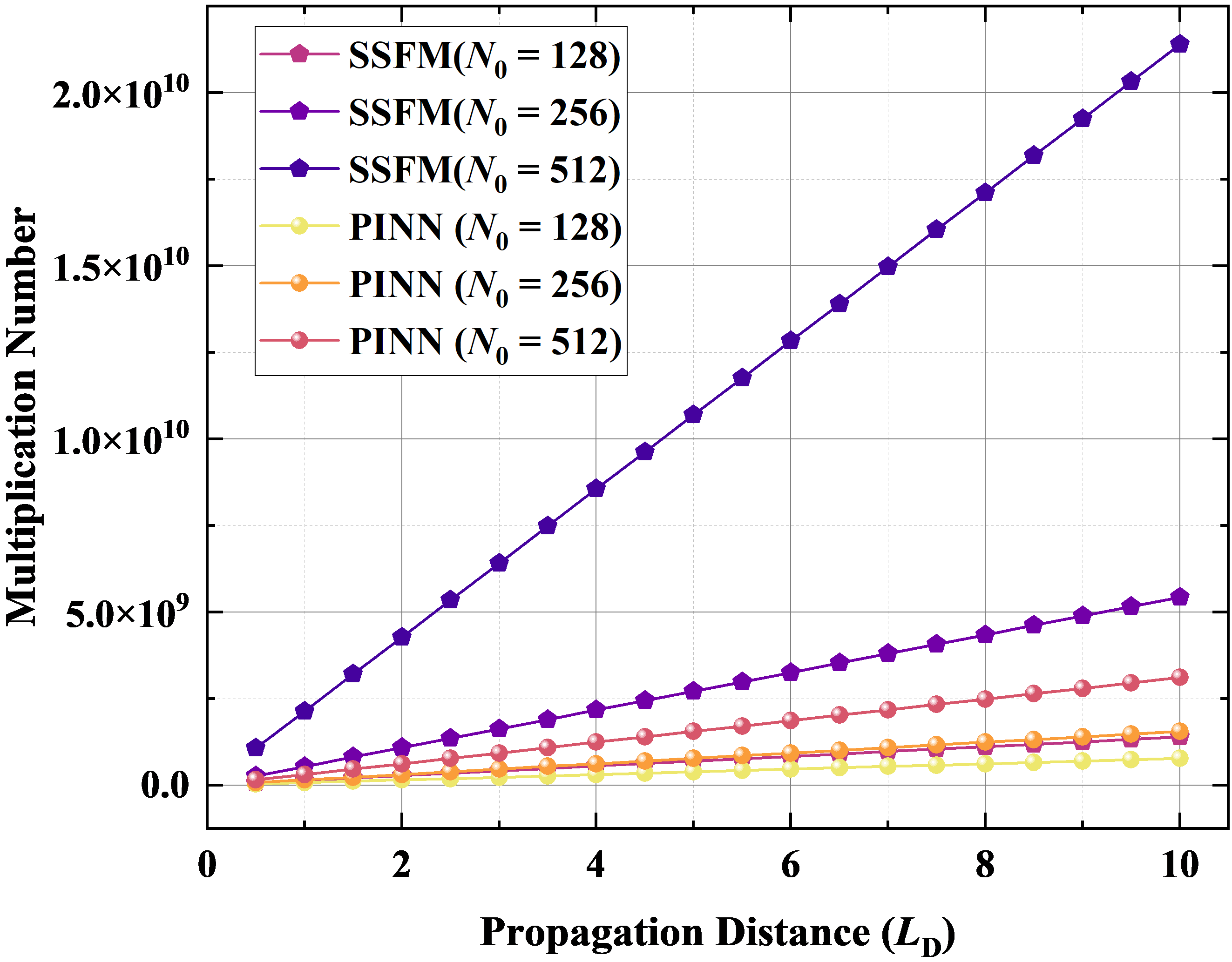}
		\setlength{\abovecaptionskip}{0pt}
    \setlength{\belowcaptionskip}{0pt}
	\caption{Comparison of the computational complexity between SSFM and PINN.}

\end{figure}

\section{Conclusion}
In summary, we have proposed a PINN-based solution to learn the nonlinear dynamics in fiber optics and carried out systematic investigation and comprehensive verification on multiple physical effects including dispersion, SPM and higher-order nonlinear effects in optical fibers. Results show the strong characterization ability of PINN when simulating pulse evolution under multiple physical effects, and special cases including optical soliton and multi-pulse propagation. The peak power of initial pulse and amplitudes of sub-pulses in the random waveform were embedded in PINN as physical parameter controllers, which made the PINN network learn the physical constraints corresponding to each scenario, thereby presenting robust generalizability in multi-scenario modeling. Furthermore, the impact of auxiliary coordinates number and physical constraints on the PINN model performance was specifically discussed and analyzed. Compared with DDNN that uses dozens of times the exact data, PINN that only considers the initial and received pulses constraints still presents higher accuracy. Moreover, PINN generally requires far fewer multiplications than SSFM for the same data volume, especially for a far longer initial pulse. Therefore, as an advanced computing technology that combines DL and physics, PINN is not only an effective PDE solver, but also a prospective technique to advance the scientific computing and automatic modeling in fiber optics.

\section*{Acknowledgements}
This work was supported in part by National Natural Science Foundation of China (No. 6217010495, 61975020), the Fund of State Key Laboratory of IPOC (BUPT) (No. IPOC2020ZT05), the Key Laboratory Fund (No. 6142104190207).

\section*{References}
% \bibliographystyle{unsrt}
%\bibliography{references}
\setlength{\parindent}{0em}
[1]	E Atlee Jackson. Perspectives of Nonlinear Dynamics: Volume 1, volume 1. CUP Archive, 1989.

[2]	Leonid P Shil’nikov. Methods of qualitative theory in nonlinear dynamics, volume 5. World Scientific, 2001.

[3]	Davood Younesian, Ali Hosseinkhani, Hassan Askari, and Ebrahim Esmailzadeh. Elastic and viscoelastic foundations: a review on linear and nonlinear vibration modeling and applications. Nonlinear Dynamics, pages 1–43, 2019.

[4]	F Calogero. Ground state of a one-dimensional n-body system. Journal of Mathematical Physics, 10(12):2197–2200, 1969.

[5]	Robert J Noll. Zernike polynomials and atmospheric turbulence. JOsA, 66(3):207–211, 1976.

[6]	Yaya Doumbia, Tushar Malica, Delphine Wolfersberger, Krassimir Panajotov, and Marc Sciamanna. Nonlinear dynamics of a laser diode with an injection of an optical frequency comb. Optics Express, 28(21):30379–30390, 2020.

[7]	Kristin M Spaulding, Darryl H Yong, Arnold D Kim, and J Nathan Kutz. Nonlinear dynamics of mode-locking optical fiber ring lasers. JOSA B, 19(5):1045–1054, 2002.

[8]	Stefan Wabnitz and Christophe Finot. Theory of parabolic pulse propagation in nonlinear dispersion-decreasing optical fiber amplifiers. JOSA B, 25(4):614–621, 2008.

[9]	Iosif Norayrovich Sisakyan and Aleksandr B Shvartsburg. Nonlinear dynamics of picosecond pulses in fiber-optic waveguides. Soviet Journal of Quantum Electronics, 14(9):1146, 1984.

[10] Atsushi Uchida. Optical communication with chaotic lasers: applications of nonlinear dynamics and synchroniza-tion. John Wiley \& Sons, 2012.

[11] Govind P Agrawal. Nonlinear fiber optics. In Nonlinear Science at the Dawn of the 21st Century, pages 195–211. Springer, 2000.

[12] RH Hardin and FD Tappert. Siam rev. Chronicle, 15:423, 1973.

[13] Robert A Fisher and William K Bischel. Numerical studies of the interplay between self-phase modulation and dispersion for intense plane-wave laser pulses. Journal of Applied Physics, 46(11):4921–4934, 1975.

[14] Bethany Lusch, J Nathan Kutz, and Steven L Brunton. Deep learning for universal linear embeddings of nonlinear dynamics. Nature communications, 9(1):1–10, 2018.

[15] Xiangjun Jin, Jie Shao, Xin Zhang, Wenwei An, and Reza Malekian. Modeling of nonlinear system based on deep learning framework. Nonlinear Dynamics, 84(3):1327–1340, 2016.

[16] Sonia Boscolo and Christophe Finot. Artificial neural networks for nonlinear pulse shaping in optical fibers. Optics \& Laser Technology, 131:106439, 2020.

[17] Sonia Boscolo, John M Dudley, and Christophe Finot. Modelling self-similar parabolic pulses in optical fibres with a neural network. Results in Optics, 3:100066, 2021.

[18] Lauri Salmela, Nikolaos Tsipinakis, Alessandro Foi, Cyril Billet, John M Dudley, and Goëry Genty. Predicting ultrafast nonlinear dynamics in fibre optics with a recurrent neural network. Nature Machine Intelligence, 3(4):344–354, 2021.

[19] Danshi Wang, Yuchen Song, Jin Li, Jun Qin, Tao Yang, Min Zhang, Xue Chen, and Anthony C Boucouvalas. Data-driven optical fiber channel modeling: a deep learning approach. Journal of Lightwave Technology, 38(17):4730–4743, 2020.

[20] Maziar Raissi, Paris Perdikaris, and George E Karniadakis. Physics-informed neural networks: A deep learning framework for solving forward and inverse problems involving nonlinear partial differential equations. Journal of Computational Physics, 378:686–707, 2019.

[21] George Em Karniadakis, Ioannis G Kevrekidis, Lu Lu, Paris Perdikaris, Sifan Wang, and Liu Yang. Physics-informed machine learning. Nature Reviews Physics, 3(6):422–440, 2021.

[22] Maziar Raissi, Alireza Yazdani, and George Em Karniadakis. Hidden fluid mechanics: Learning velocity and pressure fields from flow visualizations. Science, 367(6481):1026–1030, 2020.

[23] Dongkun Zhang, Ling Guo, and George Em Karniadakis. Learning in modal space: Solving time-dependent stochastic pdes using physics-informed neural networks. SIAM Journal on Scientific Computing, 42(2):A639–A665, 2020.

[24] Shengze Cai, Zhicheng Wang, Sifan Wang, Paris Perdikaris, and George Em Karniadakis. Physics-informed neural networks for heat transfer problems. Journal of Heat Transfer, 143(6):060801, 2021.

[25] Navid Zobeiry and Keith D Humfeld. A physics-informed machine learning approach for solving heat trans-fer equation in advanced manufacturing and engineering applications. Engineering Applications of Artificial Intelligence, 101:104232, 2021.

[26] Atilim Gunes Baydin, Barak A Pearlmutter, Alexey Andreyevich Radul, and Jeffrey Mark Siskind. Automatic differentiation in machine learning: a survey. Journal of machine learning research, 18, 2018.

[27] Arinan Dourado and Felipe AC Viana. Physics-informed neural networks for missing physics estimation in cumulative damage models: a case study in corrosion fatigue. Journal of Computing and Information Science in Engineering, 20(6):061007, 2020.

[28] Yuyao Chen, Lu Lu, George Em Karniadakis, and Luca Dal Negro. Physics-informed neural networks for inverse problems in nano-optics and metamaterials. Optics express, 28(8):11618–11633, 2020.

[29] Xiaotian Jiang, Danshi Wang, Qirui Fan, Min Zhang, Chao Lu, and Alan Pak Tao Lau. Solving the nonlinear schrödinger equation in optical fibers using physics-informed neural network. In 2021 Optical Fiber Communica-tions Conference and Exhibition (OFC), pages 1–3. IEEE, 2021.

[30] Juncai Pu, Jun Li, and Yong Chen. Solving localized wave solutions of the derivative nonlinear schrodinger equation using an improved pinn method. arXiv preprint arXiv:2101.08593, 2021.

[31] F DeMartini, CH Townes, TK Gustafson, and PL Kelley. Self-steepening of light pulses. Physical Review, 164(2):312, 1967.

[32] Keith J Blow and David Wood. Theoretical description of transient stimulated raman scattering in optical fibers. IEEE Journal of Quantum Electronics, 25(12):2665–2673, 1989.

[33] Michael Stein. Large sample properties of simulations using latin hypercube sampling. Technometrics, 29(2):143–151, 1987.

[34] David F Shanno. Conditioning of quasi-newton methods for function minimization. Mathematics of computation, 24(111):647–656, 1970.

[35] Dong C Liu and Jorge Nocedal. On the limited memory bfgs method for large scale optimization. Mathematical programming, 45(1):503–528, 1989.

[36] VI Karpman. Self-modulation of nonlinear plane waves in dispersive media. Soviet Journal of Experimental and Theoretical Physics Letters, 6:277, 1967.

[37] Antonio Napoli, Zied Maalej, Vincent AJM Sleiffer, Maxim Kuschnerov, Danish Rafique, Erik Timmers, Bernhard Spinnler, Talha Rahman, Leonardo Didier Coelho, and Norbert Hanik. Reduced complexity digital back-propagation methods for optical communication systems. Journal of lightwave technology, 32(7):1351–1362, 2014.

[38] Bernhard Spinnler. Equalizer design and complexity for digital coherent receivers. IEEE Journal of Selected Topics in Quantum Electronics, 16(5):1180–1192, 2010.

[39] Nevio Benvenuto, Giovanni Cherubini, and Stefano Tomasin. Algorithms for communications systems and their applications. John Wiley \& Sons, 2021.
\end{spacing} 
\end{document}